\documentclass[notoc,11pt,paper]{JHEP3}
\date{March 2008}
\usepackage{epsfig}
\usepackage{latexsym}
\usepackage{amssymb}
\usepackage{booktabs}
\newcommand{\be}{\begin{equation}}
\newcommand{\ee}{\end{equation}}
\newcommand{\ba}{\begin{eqnarray}}
\newcommand{\ea}{\end{eqnarray}}
\newcommand{\bi}{\begin{itemize}}
\newcommand{\ei}{\end{itemize}}
\newcommand{\ben}{\begin{enumerate}}
\newcommand{\een}{\end{enumerate}}

\newcommand{\tr}{{\rm Tr\,}}
\newcommand{\re}{\mathop{\rm Re}}

\newcommand{\nn}{\nonumber \\}

\newcommand{\half}{{\textstyle\frac{1}{2}}}

\newcommand{\N}{{\cal N}}

\newcommand{\<}{\langle}
\renewcommand{\>}{\rangle}
\newcommand{\eq}{Eq.~}

\newcommand{\fig}{Fig.~}

\newcommand{\la}{\label}

\newcommand{\txts}{\textstyle}

\newcommand{\Nt}{N_{\tau}}

\def\ov{\over}

\def\tr{{\rm tr}}

\newcommand\om{\omega}

\def\eeq{\end{equation}}
\def\half{\frac{1}{2}}
\def\Im{\mathop{\rm Im}}

\newcommand\sN{{\ensuremath{{\mathcal N}}}}
\newcommand\sO{{\ensuremath{{\mathcal O}}}}

\title{Spatial correlators in strongly coupled plasmas}

\author{
Nabil Iqbal, Harvey~B.~Meyer \\
Center for Theoretical Physics\\
Massachusetts Institute of Technology\\
Cambridge, MA 02139, U.S.A.\\
E-mail: \email{niqbal@mit.edu, meyerh@mit.edu}
}

\preprint{MIT-CTP 4058}

\abstract{
We numerically calculate the spatial correlators of the scalar and pseudoscalar
operators $F^2$ and $F\tilde F$, in SU(3) Yang-Mills theory
at zero and finite-temperature on the lattice.
We compare the results over the distances $\frac{1}{2T}<r<\frac{3}{2T}$
to the free-field prediction, to the operator-product expansion as well as to 
the strongly coupled large-$N_c$ $\sN=4$ super-Yang-Mills theory,
where results are obtained by AdS/CFT methods.
For $T_c<T<1.15T_c$, both channels exhibit stronger
spatial correlations  than in the vacuum,  
and we give an explanation for this, 
using sum-rules and the operator-product expansion.
The AdS/CFT calculation provides a semi-quantitatively successful
description of the vacuum-subtracted 
$F^2$ correlator, renormalized in the 3-loop $\overline{\rm MS}$ scheme, 
in the interval of temperatures $1.2<T/T_c<1.9$, 
while the free-field prediction has the wrong sign.
The $F\tilde F$ and $F^2$ correlators are predicted to have the same functional 
form both at weak coupling and in the strongly coupled SYM theory.
The Yang-Mills plasma does not meet that expectation below $2T_c$.
Instead we find that strong fluctuations of $F\tilde F$ 
are present at least up to that temperature.
We discuss the impact of our results on our understanding of the quark-gluon plasma.
}

\begin{document}
\section{Introduction}
Heavy ion collisions at RHIC have revealed properties of the 
quark gluon plasma that had not been widely anticipated
(see~\cite{Muller:2008zzm} for an introduction).
The ability of the produced matter to flow with little dissipation
and to strongly quench energetic jets seemed to disfavor 
a description of the matter in terms of weakly interacting 
quarks and gluons. On the other hand, 
the constituent quark number scaling
of the measured elliptic flow coefficient 
(see for instance \cite{Adare:2006ti})
suggests that it is particles with the quantum numbers of quarks
that are flowing in the expanding fireball.

One of the central questions is thus whether 
the quark-gluon plasma at temperatures within reach of 
heavy-ion collisions is better described in 
a weak coupling expansion or whether a radically different 
computational scheme is more appropriate.
The answer to the question could depend on the quantity,
in which case it would be even more difficult 
to form a mental picture of the plasma.

The strong elliptic flow and jet quenching observed 
in heavy ion collisions point to very strong interactions 
among the constituents of the plasma.
Indeed the quantities most sensitive
to interactions appear to be the dynamical ones, 
such as the shear viscosity $\eta$ in units of the entropy density $s$,
which varies like $\alpha_s^{-2}$ 
between order unity and $+\infty$ with the coupling.
Such dynamic properties of the plasma remain a challenge
for lattice calculations (see~\cite{Meyer:2009jp} for a review). 
In the shear channel the most accessible transport property
is 
$\int_0^\Omega d\omega \rho(\omega)/\omega$ with $\Omega$ of order $T$,
where $\rho$ is the spectral density.
For a weakly coupled system obeying the $f$-sum rule~\cite{Teaney:2006nc}, 
this provides a measure of the mean square velocity $v_{\bf p}^2$ 
of the quasiparticles responsible
for the transverse transport of momentum.
A value much below unity would rule out the possibility of 
these quasiparticles being light quarks or gluons.

Static quantities on the other hand, while often providing
a less clear-cut test of the importance of interactions,
are directly accessible in the Euclidean formulation of the theory.
The thermodynamic potentials, for instance,
have remained a challenge for perturbative 
methods~\cite{Kajantie:2002wa,Hietanen:2008tv},
even though certain resummation schemes lead 
to more stable predictions~\cite{Blaizot:2003iq,Blaizot:2003tw}.
The example of strongly coupled ${\cal N}=4 $  super-Yang-Mills (SYM) theory, 
where the entropy density is only reduced by a factor 3/4
with respect to the non-interacting case~\cite{Gubser:1996de}, shows 
that only a highly accurate agreement of the weak coupling
expansion and non-perturbative 
lattice data can warrant the  conclusion
that the plasma is dominated by weakly coupled 
quark and gluon quasiparticles.

Other static quantities on the other hand appear to be 
quite well described by weak coupling techniques. 
A convincing  example is the spatial string tension, for which
the dimensional reduction program works well~\cite{Laine:2005ai,Alanen:2009ej}.
As another example, the fluctuations of quark numbers~\cite{Cheng:2008zh}
appear to approach remarkably early the Stefan-Boltzmann limit. 
Recently the expectation values of other twist-two operators
(besides the energy-momentum tensor)
have been proposed~\cite{Iancu:2009zz} as diagnostic tools for the 
effective strength of interactions.

In this paper we calculate non-perturbatively spatial correlators of
two dimension-four operators, the trace anomaly $\theta(x)$
and the topological charge density $q(x)$ in the SU(3) gauge theory.
The range of distances covered by the calculation is 
$\frac{1}{2T}<r<\frac{3}{2T}$.
These are also static quantities that are directly calculable 
on the lattice. Furthermore, quite a lot is known about
these correlators in the QCD vacuum, going back to the original
QCD sum rules studies~\cite{Novikov:1979va,Novikov:1980uj}. 
And thirdly, they are computable 
in the large-$N_c$, strongly coupled ${\cal N}=4$ SYM theory 
by AdS/CFT methods~\cite{Maldacena:1997re,Gubser:1998bc,Witten:1998zw}. 
Thus we have the possibility to compare the lattice data to 
two `caricatures' of the plasma, one being non-interacting gluons and 
the other being a very strongly coupled non-Abelian plasma.
As we shall see, once the vacuum contribution has been subtracted
these two caricatures lead to qualitatively different 
predictions for the correlators.

Parallel to the question of the weak- or strong-coupling nature
of the quark-gluon plasma, lies the question of how 
similar non-Abelian relativistic plasmas are.
This constitutes a very interesting question in itself.
In addition, the possiblity to compute real-time quantities 
in strongly coupled theories amenable to AdS/CFT computations
and ``port'' them to QCD provides a strong phenomenological motivation.
The best known examples of this strategy are the shear viscosity 
to entropy density ratio~\cite{Kovtun:2004de} 
and the jet quenching parameter $\hat q$ calculations~\cite{Liu:2006ug}.
As evidence in favor of the strategy, 
the authors of~\cite{Bak:2007fk} conclude that the overall agreement
of the screening spectra of QCD and the ${\cal N}=4$ SYM theory
is rather good, although the low-lying screening masses are overall 
a factor 1.9 or so larger in the strongly coupled SYM theory.
They therefore suggest that the QCD plasma around $2T_c$ is 
most similar to the ${\cal N}=4$ SYM plasma at an intermediate
value of the 't~Hooft coupling $\lambda$. In order to 
find the effective coupling which leads to the best match
(defined by a set of physical quantities) between the theories
therefore requires knowing the properties of the ${\cal N}=4$ SYM plasma
at intermediate values of the coupling,  presumably 
as hard a problem as determining those of the QCD plasma.
However, in the SYM theory one has the advantage of being 
able to expand the observables in $\lambda$ and in $1/\lambda$,
opening the possibility to interpolate to 
intermediate couplings~\cite{Bak:2007fk}.

The outline of this paper is as follows. In section (\ref{sec:defs}) we 
define the relevant operators and their correlators, 
and give the basic free-field theoretic predictions. 
Section (\ref{sec:sym}) contains the AdS/CFT calculation of the 
same correlators in the strongly coupled ${\cal N}=4$ SYM theory.
In section (\ref{sec:lat}) we describe the lattice calculation of these 
correlators, including a new way to normalize the 
topological charge density for on-shell correlation functions.
The results are compared to weak- and strong-coupling theoretical 
predictions. 
Section (\ref{sec:coupling}) discusses what values of the  't~Hooft coupling 
best match the gluonic and the SYM plasma.
We finish with a summary of the lessons learnt and an outlook in 
section (\ref{sec:disc}).

\section{Definitions and theoretical predictions\la{sec:defs}}
In this section and in the following, we use Euclidean conventions, 
since the calculation of correlators will be performed on the lattice.
We consider the SU($N_c$) gauge theory without matter fields,
\be
S_{\rm E} =  \frac{-1}{2g^2}\int d^4x \,\tr\{F_{\mu\nu}(x)F_{\mu\nu}(x)\}\,.
\ee
We focus on two operators in this paper. 
The first is the (anomalous) trace of the energy-momentum tensor,
\ba
\theta(x)\equiv T_{\mu\mu}(x) 
&=& {\txts\frac{\beta(g)}{2g}} ~ F_{\rho\sigma}^a  F_{\rho\sigma}^a\,,
\qquad
\beta(g) =  -bg^3+\dots,\qquad  b={\txts\frac{11N_c}{3(4\pi)^{2}}}\,.
\ea
The second operator is the topological charge density. It is defined as 
\be
q(x) = \frac{-1}{32\pi^2}\epsilon_{\mu\nu\rho\sigma} 
\tr\{F_{\mu\nu}(x)F_{\rho\sigma}(x)\}
= \frac{g^2}{32\pi^2}  F^a_{\mu\nu}(x) \tilde F^a_{\mu\nu}(x)
\ee
where $F_{\mu\nu}=g\,F_{\mu\nu}^a t^a $, $\tr\{t^at^b \}=-\half\delta_{ab}$ 
and $\tilde F^a_{\mu\nu}(x)\equiv 
\half\epsilon_{\mu\nu\rho\sigma}F^a_{\rho\sigma}(x)$.
The normalization is chosen such that 
the value of $Q=\int d^4x\, q(x)$ on a self-dual configuration is an integer.
For later use we also introduce the operator 
\be
\theta_{00}(x) = \frac{1}{4}F_{ij}^aF_{ij}^a-\frac{1}{2}F_{0i}F_{0i}\,.
\ee
In the thermodynamic limit, $\<\theta_{00}\>_{T-0}=e+p$ while 
$\<\theta\>_{T-0} = e-3p$, where $e$ and $p$ are the energy density 
and pressure respectively and the subscript $T-0$ means that the difference
of the thermal expectation value and the vacuum expectation value is taken.
For ${\cal O}= \theta$ or $q$, we will consider the static connected 
correlators at finite temperature $T\equiv 1/L_0$,
\be
C_{\cal O O}(r,T) \equiv \<{\cal O}(0,{\bf r}) {\cal O}(0)\>_c
= \frac{1}{Z(L_0)} {\rm Tr}\left[e^{-L_0 H}\, \hat {\cal O}(0,{\bf r})\,\hat {\cal O}(0)\right]
 - \left(\frac{1}{Z(L_0)} {\rm Tr}\left[e^{-L_0 H}\, \hat {\cal O}(0)\right]\right)^2\,.
\la{eq:statcor}
\ee
Often, to emphasize the thermal effects on the correlator, we will subtract 
the zero-temperature correlator,
\be
G_{\cal O O}(r,T) \equiv C_{\cal O O}(r,T)  - C_{\cal O O}(r,0) \,.
\ee
If one expresses the traces of \eq(\ref{eq:statcor}) in a basis of 
energy eigenstates, this subtraction has the effect of removing 
the vacuum contribution.
\subsection{Short-distance behavior}
In this section we review the available perturbative results for the
correlators (\ref{eq:statcor}) as well as our knowledge of their 
long-distance behavior.
\subsubsection{Zero temperature}
The  two-point functions of the trace anomaly and the 
topological charge density are to leading order
\be
(8\pi b\alpha_s)^{-2}\<\theta(x)\theta(0)\>_{\rm 1\,loop}  =
-\left(\frac{2\pi}{\alpha_s}\right)^2\, \<q(x)\,q(0)\>_{\rm 1\,loop} = 
\frac{3d_A}{\pi^4(x^2)^4}\,,
\la{eq:ththPT}
\ee
where $d_A=N_c^2-1$ is the number of gluons.
The correlators were calculated to two-loop order in~\cite{Kataev:1981gr},
but we will not exploit that result.

\subsubsection{Finite temperature}
The  two-point functions of the trace anomaly and the 
topological charge density  to leading order read~\cite{Meyer:2008dt}
\be
(8\pi b\alpha_s)^{-2}\<\theta(x)\theta(0)\>_{\rm 1L}  =
-{\txts\left(\frac{2\pi}{\alpha_s}\right)^2}\, \<q(x)\,q(0)\>_{\rm 1L} = 
 \frac{d_A}{\pi^4}\sum_{m,n\in {\bf Z}}
\left(4 \frac{(x_{[m]}\cdot x_{[n]})^2} {x_{[m]}^2x_{[n]}^2} -1\right)
\frac{1}{(x_{[m]}^2 \,x_{[n]}^2)^2 }\,,
\la{eq:free}
\ee
where $x_{[n]} \equiv x+n L_0 \hat e_0$.
Their operator-product expansion (OPE) reads~\cite{Novikov:1979va,Meyer:2008dt}
\ba
(8\pi b\alpha_s)^{-2}\<\theta(x)\theta(0)\>
= -{\txts\left(\frac{2\pi}{\alpha_s}\right)^2}\, \<q(x)\,q(0)\> 
+\dots
= \frac{3d_A}{\pi^4 r^8}
- \frac{1}{3\pi^2} \frac{\<\theta_{00}\>}{r^4}
- \frac{1}{2\pi^2} \frac{\<\theta\>}{r^4}
+\dots
\la{eq:OPEth}
\ea
and the dots refer to ${\rm O}({\txts\frac{1}{r^2}})$ terms.
In fact, the OPE of the two correlation functions appearing in \eq(\ref{eq:OPEth}),
treating the Wilson coefficients to leading order in perturbation theory,
differ only  at O(${r^0}$) if one restricts the terms on the 
right-hand side to operators whose vacuum expectation value 
does not vanish~\cite{Novikov:1980uj}.
Since the $r^{-8}$ term cancels exactly when taking the difference of two temperatures
and $\<\theta_{00}\>_{T-0}\geq0$, $\<\theta\>_{T-0}\geq0$,
\eq(\ref{eq:OPEth}) implies that 
the gluon plasma is always more screening than the vacuum of the theory
at sufficiently small distances.

\subsubsection{Spectral functions}
The free spectral functions for the trace anomaly and the 
topological charge density are~\cite{Meyer:2008gt}
\ba
(8\pi b\alpha_s)^{-2} \rho_{\theta,\theta}(\omega,q,T) &=&
- {\txts\left(\frac{2\pi}{\alpha_s}\right)^2}\, \rho_{qq}(\omega,q,T) 
= \frac{d_A}{(8\pi)^2} ~(\omega^2-q^2)^2 ~~  {\cal I}([1], \omega, q, T)\,,
\nn
{\cal I}([1], \omega, q, T) &=&
-\frac{\omega}{q} \theta(q-\omega)
\,+\, \frac{2T}{q}\,\log\frac{\sinh(\omega+q)/4T}{\sinh{|\omega-q|/4T}}\,.
\la{eq:sfelem}
\ea
The spectral functions are related to the static correlators by 
Fourier transformation,
\be
\<{\cal O}(0,{\bf r})\, {\cal O}(0)\>
= \lim_{\epsilon\to0}\int \frac{d^3{\bf q}}{(2\pi)^3}e^{i{\bf q\cdot r}}
\int_0^\infty d\omega e^{-\epsilon \omega}
\frac{\rho(\omega,{\bf q},T)}{\tanh\omega/2T}\,.
\ee
The parameter $\epsilon$ serves to regulate the integral over
frequencies at large $\omega$, which would otherwise diverge.
We will come back to this regulator in section 4.

\subsection{Long-distance behavior}
At long distances, the vacuum correlators of $\theta$ and $q$ are dominated 
by the scalar and pseudoscalar glueballs respectively.
The most recent lattice results for their masses are
$r_0M_{0^{++}} = 3.96(5)$~\cite{Meyer:2008tr}
or 4.16(11)(4)~\cite{Chen:2005mg}
and $r_0M_{0^{-+}} = 5.93(16)$~\cite{Meyer:2004jc}
or 6.25(6)(6)~\cite{Chen:2005mg}, 
where $r_0\simeq 0.5$fm is the Sommer reference scale~\cite{Sommer:1993ce}.
The coupling of these states to the local operators $\theta$ and $q$, 
$s\equiv\<{\rm vac}|\theta|0^{++}\>$ and $p\equiv\<{\rm vac}|q|0^{-+}\>$,
have also been calculated recently~\cite{Chen:2005mg,Meyer:2008tr}.

The screening masses, which determine the asymptotic 
exponential fall-off of the finite-$T$ correlators, are also known to some extent.
The operators $\theta$ and $q$ belong to irreducible representations (irreps)
of the SO(3) rotation group, $\times$ parity and $\times$ charge conjugation.
At finite temperature, the symmetry group of a `$z$-slize' (for states
at rest, $p_x=p_y=0$ and given $\omega_n$, the discrete momentum in the 
direction of length $1/T$) is reduced to R$\times$SO(2)$\times P_2\times C$, where
R is the Euclidean-time reflection and
$P_2$ is the reflection inside an $xy$ plane ($(x,y)\to(x,-y)$).
In general, an operator forming an irrep of the zero-temperature theory
gets decomposed into several irreps of this reduced symmetry group.
In our case, $\theta$ and $q$ simply become the scalar and pseudoscalar
representations of the $z$-slice symmetry group. The former is invariant 
under all the symmetries of the $z$-slice; the latter is too, except
for being odd under the  R and $P_2$ operations. 
On the lattice, these irreps are generically further
reduced to crystallographic irreps. In our case they
are labelled  $A_1^{++}$ and $A_1^{-+}$~\cite{Datta:1998eb}.
A recent result for the mass gap in the scalar sector is $m_{A_1^{++}}(T)/T= 2.62(16)$,
2.83(16) and 2.88(10) respectively at $1.24T_c$, 
$1.65T_c$ and $2.20T_c$~\cite{Meyer:2008dt}.
Datta and Gupta find $m_{A_1^{-+}}(T)/T= 6.32(15)$ both at about $1.5T_c$ 
and $2.0T_c$~\cite{Datta:1998eb}.
So the asymptotic screening is much stronger in the pseudoscalar sector
than in the scalar sector. This ordering persists at all temperatures
according to a recent next-to-leading order perturbative analysis,
in spite of a change in the nature of the lightest scalar state~\cite{Laine:2009dh}.

\section{AdS/CFT Results for $\sN = 4$ Plasma} \label{sec:sym}
We now turn to the calculation of spatial correlators
in a maximally supersymmetric strongly 
coupled plasma using  gauge-gravity 
duality~\cite{Maldacena:1997re,Gubser:1998bc,Witten:1998zw}. 
We will find that the thermal correlators in the $\sN = 4$ 
plasma are identical for the operators $F^2$ and $F\tilde{F}$, 
as each of these operators is dual to a simple massless scalar field. 
It is interesting that the two correlators also coincide at weak
coupling, where they are given by a two-gluon exchange diagram,
and therefore coincide with the pure Yang-Mills result (\ref{eq:free}).

These 
correlators have been previously studied in momentum space 
in \cite{Hartnoll:2005ju}. See 
also~\cite{Kovtun:2006pf, Teaney:2006nc} for discussion of 
finite temperature stress tensor and R-current correlators 
in momentum space.
An outline of the computation is:
\ben
\item letting $\sO$ denote either $F^2$ or $F\tilde{F}$, 
we note that the field theory operator $\sO$ is dual to 
a massless bulk scalar field $\phi$. For $F^2$ this field 
$\phi$ is exactly the type IIB dilaton, whereas for 
$F\tilde{F}$ $\phi$ it is the Ramond-Ramond axion $C_0$. 
\item We then compute the spectral density $\rho(\omega, k)$ 
for the operator $\sO$ using finite-temperature AdS/CFT. 
This involves numerically solving the bulk equations of 
motion for $\phi$ in a black brane geometry.
\item Finally, we Fourier transform this spectral density 
to obtain the Euclidean correlator in position space.
\een
Each of these steps is explained in more detail below. 
Throughout we will expand fields on each constant-radius 
slice in Fourier space, 
$\phi \sim \phi(r) e^{-i\omega t + i k z}$. 
We take the spatial momentum to be in the $z$ direction. 
With an eye on numerical evaluation,
we will often work with dimensionless momenta 
and positions, which we denote with an overbar:
\be
\bar{\omega} = \frac{\omega}{2\pi T} \qquad \bar{x} = 2\pi T x
\ee 
The relevant black brane metric for $\sN = 4$ SYM at 
finite temperature on $\mathbb{R}^{3,1}$ can be written
\be
ds^2 = (\pi R T)^2 r^2\left[-\left(1-\frac{1}{r^4}\right)dt^2 
+ d\vec{x}_3^2\right] + \frac{1}{1-\frac{1}{r^4}}
\frac{dr^2 R^2}{r^2}, \label{bhmetric}
\ee
where $R$ is the radius of the bulk AdS space and $T$ 
the temperature of the black brane, with the horizon 
at $r = 1$ and the AdS boundary at $r \to \infty$. 

\subsection{Flow Equation and Numerical Evaluation}
We now let $\sO$ denote either $F^2$ or $F\tilde{F}$. 
In both cases the relevant bulk action for the field 
dual to $\sO$ is simply that of a massless scalar
\be
S = -\frac{1}{2\alpha}\int d^5 x\sqrt{-g}(\nabla\phi)^2,
\ee
For these operators supersymmetry guarantees that the 
vacuum two-point function is independent of the 
coupling~\cite{Hartnoll:2005ju}, and thus the normalization $\alpha$ 
can be conveniently determined by demanding that this 
correlator as computed from gravity agrees with the 
free-field expressions (\ref{eq:ththPT}). We find for both
\be
\frac{1}{2\alpha_{F^2}} = \frac{1}{2\alpha_{F\tilde{F}}}
 = \frac{N^2}{\pi^2 R^3} 
\ee
The spectral density $\rho$ is proportional to the 
imaginary part of the finite temperature retarded 
correlator, $\rho = -\frac{1}{\pi}\Im(G_R)$. 
An extensive literature exists on the evaluation 
of this quantity from
 AdS/CFT~\cite{Son:2002sd, vanRees:2009rw, 
Skenderis:2008dg, Skenderis:2008dh}. We will use the 
flow formalism developed in \cite{Iqbal:2008by}, 
which we briefly review here: consider the function 
$\chi(r,k)$, defined as
\be
\chi(r,k) \equiv \frac{\Pi(r,k)}{i\om\phi}
 \qquad \Pi(r,k) = -\frac{1}{\alpha}\sqrt{-g}g^{rr}\partial_r\phi.
\ee
Here $\Pi(r,k)$ is the momentum conjugate to 
the bulk field $\phi(r,k)$. The bulk equation 
of motion for $\phi$ implies that the $\chi(r,k)$ 
obeys (on any metric) the first-order flow equation
\be
\partial_r \chi = i\omega \sqrt{\frac{g_{rr}}{g_{tt}}}
\left[\frac{\chi^2}{\Sigma_\phi} - \Sigma_\phi
\left(1 - \frac{k^2 g^{zz}}{\omega^2 g^{tt}}\right)\right]
 \qquad \Sigma_\phi = \frac{1}{\alpha}\sqrt{\frac{-g}{g_{rr}g_{tt}}}
\ee
Furthermore it follows from real-time AdS/CFT~\cite{Iqbal:2009fd}
 that the retarded correlator $G_R$ in the dual field theory
 is obtained from the boundary value of $\chi(r,k)$:
\be
G_R(k) = -\lim_{r\to\infty}i\omega\chi(r,k) \qquad \to
 \qquad \mathrm{Im}[G_R(k)] = -\lim_{r\to\infty}\omega
 \mathrm{Re}[\chi(r,k)].
\ee
The initial conditions at the horizon $r = 1$ are
 fixed by the infalling wave condition to be
 $\chi(r = 1) = \Sigma_\phi(r = 1)$. 

We now plug in the metric (\ref{bhmetric}) and 
define a dimensionless function $\tilde{\chi}$ by $\tilde{\chi}
 = \frac{\alpha\chi}{(\pi R T)^3}$. We obtain the flow equation
\be
\partial_r\tilde{\chi} = 
\frac{2 i \bar{\omega}}{r^2\left(1 - \frac{1}{r^4}\right)}
\left[{\tilde{\chi}^2 \ov r^3} - r^3
\left[1-\frac{\bar{k}^2}{\bar{\omega}^2}
\left(1-\frac{1}{r^4}\right)\right]\right]. \label{floweqn}
\ee
This equation must now be integrated from 
$\tilde{\chi}(z = 1)=1$ to the AdS boundary at 
$z = \infty$, where it determines the AdS/CFT response. 
Some technical details on the numerical integration 
are given in Appendix \ref{app:numerics}. 

\subsection{Fourier Transform}
To obtain a Euclidean correlator from the spectral 
density, we use the identity \cite{Kapusta:2006pm}.
\be
G_E(0;\tau,x) =  \int_0^\infty d\omega \int
 \frac{d^3 \vec{k}}{(2\pi)^3} \;\rho(\omega,\vec{k})
 \frac{\cosh\left(\omega
\left(\tau - \frac{\beta}{2}\right)\right)}
{\sinh(\frac{\omega\beta}{2})}e^{i\vec{k}\cdot\vec{x}}
\ee
Assuming rotational symmetry in the spatial directions
(i.e. $\rho(\omega, \vec{k}) = \rho(\omega, |\vec{k}|)$)
to perform the angular integral and switching to
dimensionless momenta, we obtain
\be
G_E(0;\bar{\tau},|\bar{x}|) = 8\pi^2 T^4
 \int_0^{\infty}d\bar{\omega}\int_0^{\infty}
d|\bar{k}|\;\rho(\bar{\omega},\bar{k}) \sin(|\bar{k}||\bar{x}|)
 \frac{\bar{k}}{\bar{x}}\frac{\cosh\left(\bar{\omega}\left(\bar{\tau}
 - \pi\right)\right)}{\sinh(\bar{\omega}\pi)}
\ee
We note at this point that we are primarily interested
in equal time correlators, i.e. those for which $\tau$ 
is $0$ in the equation above. 
However, at small time separations the nonzero value
of $\tau$ acts like a UV cutoff on high-frequency modes,
suppressing them as $e^{-\omega\tau}$. To achieve arbitrarily 
small $\tau$ we would need to know $\rho$ at arbitrarily high
$\omega$, whereas numerically we are necessarily limited to finite
$\omega$\footnote{Note this is an advantage to using
the real-time formalism described here, as there would be
no such exponential suppression of high frequency modes
if we were to compute the position space correlator by
summing the Euclidean momentum space correlator over
Matsubara modes.}. 
Since the position-space, equal-time correlator is finite,
one could repeat the calculation with
several $\tau$ values and extrapolate to $\tau=0$.
Here we however restrict ourselves to using a small, finite 
`regulator' $\tau\ll r$. Since our goal is to compare the 
correlators to those computed on the lattice, where 
the region of very small $r$ is in any case 
problematic due to discretization errors,
this approach will prove sufficient.

As a check on the numerical Fourier transforms themselves,
we compute the corresponding Fourier transform 
in the free theory at finite
temperature starting from the analytic expression for 
the spectral density \eq(\ref{eq:sfelem}); in this case an
analytic expression also exists directly in position space,
\eq(\ref{eq:free}), providing a non-trivial
check on the accuracy of the Fourier transform. In both
cases we subtract the zero-temperature contribution,
which, as mentioned above, is independent of the coupling.
The step sizes in $\bar\omega$ and $\bar k$ are both 0.1,
and the chosen range of integration is 0 to 20.
The result of this numerical integration is shown 
in \fig\ref{fig:justone}, where we have fixed 
$\tau = \frac{1}{2\pi T}$.
With $x$ ranging over values much larger than $\tau$,
the Figure shows the departure of the Fourier-transformed 
correlator from the direct evaluation of the $x$-space
expression~(\ref{eq:free}). The achieved accuracy is sufficient 
for our purposes. The largest discrepancy occurs at short distances,
where the sensitivity to the discretization step and the 
finite range of $\omega$ is greatest.

To illustrate the dependence on the regulator $\tau$, 
we compare the correlator $C_{\sO\sO}(r,\tau,T) - C_{\sO\sO}(r,\tau,0) $
for $\tau=1/2\pi T$
with $G_{\sO\sO}(r,T)$  in the free case (\fig\ref{fig:tautest}).
We see that for $r> 5/2\pi T$, the correlators coincide to 
the accuracy needed for our purposes.


\section{Correlators on the lattice\la{sec:lat}}
In this section we describe the lattice setup and the 
numerical results obtained by Monte-Carlo simulations.
We employ the Wilson action~\cite{Wilson:1974sk},
\be
 S_{\rm g} =  \frac{1}{g_0^2} \sum_{x,\mu\neq\nu} \tr\{1-P_{\mu\nu}(x)\}\,,
\la{eq:Sg}
\ee
where the `plaquette' $P_{\mu\nu}$ is the product of four link 
variables $U_\mu(x)$ around an elementary cell in the $(\mu,\nu)$
plane.

As a simulation algorithm, we use the standard combination of heatbath and
over-relaxation~\cite{Creutz:1980zw,Cabibbo:1982zn,Kennedy:1985nu,Fabricius:1984wp}
sweeps for the update in a ratio increasing from
3 to 5  as the lattice spacing is decreased.
The overall number of sweeps between measurements was typically
between 4 to 12.

\subsection{Discretization and normalization of the operators}
A choice has to be made for the discretization of the 
operators $\theta(x)$ and $q(x)$. 
Here we use the specific discretization
\ba
\theta_L(x) & \equiv  &
-\chi_s(g_0) \frac{dg_0^{-2}}{d\log a}
~\half\sum_{\mu,\nu} \re\tr
\Big[\widehat F_{\mu\nu}(x)\widehat F_{\mu\nu}(x)\Big]\,,
\la{eq:Th}
\\
q_L(x) &\equiv &
\frac{-Z_q(g_0)}{32\pi^2}\epsilon_{\mu\nu\rho\sigma}
\tr\Big[\widehat F_{\mu\nu}(x)\widehat F_{\rho\sigma}(x)\Big]\,,
\la{eq:Q}
\ea
where the (antihermitian) 
`clover' discretization of the field-strength tensor
$\widehat F_{\mu\nu}(x)$ is defined in terms of the 
link variables in~\cite{Luscher:1996sc}.
In this work we feed in `HYP smeared' link variables~\cite{Hasenfratz:2001hp}
into the definition of $\widehat F_{\mu\nu}(x)$.
The name stems from the fact that the 
elementary link variable is replaced by an average
of Wilson lines which remain in the adjacent elementary 
hypercubes. We kept the original parameters~\cite{Hasenfratz:2001hp}
and used the projection onto the SU(3) group
of the Wilson-line average described in~\cite{DellaMorte:2005yc}.

At the quantum level,
the normalization of these lattice operators 
differs from the naive normalization.
Indeed, even though the anomalous dimension of these
operators  vanishes, a finite renormalization of the operators
survives, which has to be compensated for
in order to ensure that their on-shell
matrix elements approach their continuum limit with O($a^2$) 
corrections.

In  \eq(\ref{eq:Th}), $\frac{dg_0^{-2}}{d\log a}$ is the lattice 
beta-function which describes by how much the lattice spacing shrinks
when the bare coupling is reduced. While asymptotically it 
is governed by the first two universal beta-function coefficients, 
we work in the region $g_0^2\sim 1$ and therefore employ 
a non-perturbatively determined beta-function. 
Specifically we use the quantity $r_0/a$ 
(the Sommer reference scale) as a function of 
$g_0^2$, as computed in~\cite{Necco:2001xg} and 
parametrized in the appendix of~\cite{Durr:2006ky}.
By taking one derivative of the parametrization, we obtain 
$\frac{dg_0^{-2}}{d\log a}$.

The trace anomaly in our chosen discretization 
still requires the additional normalization factor $\chi_s(g_0)$.
The latter is fixed by 
calibrating against the `canonical' discretization of $\theta(x)$.
This discretization $\theta_L'$
arises from differentiating the lattice action (\ref{eq:Sg}) 
with respect to the bare coupling,
\be
a^4\theta_L'(x)
=   \frac{dg_0^{-2}}{d\log a} 
\sum_{\mu,\nu} \re\tr   \Big[ 1 - P_{\mu\nu}(x)  \Big].
\ee
By requiring that $e-3p$ be independent of the discretization,
i.e. $\<\theta_L(x)\>_T-\<\theta_L(x)\>_0$ 
be equal to $\<\theta_L'(x)\>_T-\<\theta_L'(x)\>_0$,
we determine $\chi_s(g_0)$.
Here we choose $\Nt\equiv L_0/a=6$ to do this matching.
The results are well parametrized by
\be
\chi_s(g_0) = \frac{0.3257 }{ 1 - 0.7659 g_0^2},\quad
5.90\leq \beta\leq 6.72.
\ee
The error varies from 0.004 at $\beta=6.0$ to 0.008 at $\beta=6.72$.
We have not investigated systematically the dependence 
of $\chi_s$ on the value of $\Nt$ used for the matching.
This uncertainty is not included in the above error estimates,
and requires repeating the matching calculation at $\Nt=8$ or $12$.
For our purposes in this work, this uncertainty will not prevent 
us from drawing conclusions when comparing the trace anomaly correlator
to theoretical predictions.

\subsubsection{Normalization of the topological charge density}
The procedure we use to normalize $q_L(x)$ is somewhat new 
to our knowledge, and therefore we describe it in some detail.
We again exploit the fact that there exists a discretization
$q_L'(x)$ for which the normalization is known exactly.
This is the definition based on the overlap operator~\cite{Neuberger:1997fp}
which satisfies the Ginsparg-Wilson relation~\cite{Ginsparg:1981bj}.
Indeed, $Q_L'(x)=\sum_x q_L'(x)$ is then guaranteed to be 
an integer on every gauge field configuration, 
because it counts the difference of the number of right-handed
and left-handed zero-modes of the Dirac operator~\cite{Hasenfratz:1998ri}.

We normalize our discretization of $q(x)$ by matching 
the value of its vacuum two-point function 
with the same correlator obtained with the overlap-fermion-based
discretization of $q(x)$. The latter correlator was obtained 
in~\cite{Horvath:2005cv}, and we use the numerical data of that 
article to fix the normalization of our discretization.
More precisely, the quantity matched is $r^8\,C_{qq}(r,0)$;
this removes the largest part of the uncertainty in the lattice spacing
in physical units.

Specifically, we choose the matching distance to be $\bar r/r_0\simeq0.68$,
and use the data of~\cite{Horvath:2005cv} on the finest lattice 
(${\cal E}_3$ ensemble at $a=0.082$fm). The matching distance $\bar r$ 
was chosen such that on this lattice, $\bar r /a = 3\sqrt{2}$.
The lattice spacing in~\cite{Horvath:2005cv} is specified by the string tension,
and we have used the factor $r_0\sqrt\sigma = 1.1611(95)$ 
(based on the compilation~\cite{Meyer:2008tr})
to convert physical distances in units of $r_0$.

Given our normalization strategy, it is 
convenient to split the normalization factor of the operator 
into two parts:
\be
Z_q(\beta) = Z_q(\beta_{\rm ref}) \cdot \chi(\beta).
\ee
where we choose $\beta_{\rm ref}=6.2822$.
The advantage of this separation is that only $ Z_q(\beta_{\rm ref})$
depends on the overlap-based correlator. 
The latter has been calculated to much lower statistics 
than our computationally cheaper correlator. We find 
\be
 Z_q(\beta_{\rm ref}) = 1.55(8)\,,
\ee
where the uncertainty comes almost entirely from the overlap data.
We then obtain, by matching the correlator at other values of $\beta$,
\ba
\chi(5.903)&=& 1.25(4) \qquad r_{\rm match}/r_0=0.94\\
\chi(6.018)&=& 1.125(24) \qquad r_{\rm match}/r_0=0.77\\
\chi(6.200)&=& 1.030(15) \qquad r_{\rm match}/r_0=0.68 \\
\chi(6.408)&=& 0.959(20). \qquad r_{\rm match}/r_0= 0.60\,\\
\chi(6.720)&=& 0.958(50). \qquad r_{\rm match}/r_0= 0.60\,.
\ea
When necessary, 
we use linear interpolations in $r$ of the function $r^8\,C_{qq}(r,0)$
to match different lattice spacings. These $\chi$ factors are well 
parametrized as 
\be
\chi(\beta)= \frac{0.8112-0.7388 g_0^2}{1-0.9364 g_0^2},\qquad \beta=6/g_0^2\,.
\ee
The absolute error is about 0.041 at $\beta=5.9$,  0.014 at $\beta=6.1$,
0.010 at $\beta=6.4$ and 0.026 at $\beta=6.7$.

\subsection{The vacuum correlators ($T=0$)}
The vacuum correlators of the trace anomaly and the 
topological charge density (multiplied by $r^8$)
are displayed in \fig\ref{fig:edotb-h}.
The overall normalization uncertainty coming from
$\chi_s(g_0)$ and 
$Z_q(\beta)$ are not included in the error bars on the picture;
they are given above. Only data points with $r/a\geq 4$ 
are displayed.
We have measured the correlator along the lattice axes (1,0,0), 
(1,1,0) and (1,1,1) and averaged the results over all equivalent directions.
Some of the raw data is given in Tables~(\ref{tab:T=0}) and (\ref{tab:T=0b}).

The plots also show the perturbative prediction~(\ref{eq:ththPT}) 
at small distances. The three lines give an 
indication of the renormalization scale uncertainty: they correspond to
$\mu=\frac{\pi}{2r},\frac{\pi}{r},\frac{2\pi}{r}$. For that purpose we have used the 
result $\Lambda^{(N_{\rm f}=0)}_{\overline{\rm MS}}r_0=0.602(48)$~\cite{Capitani:1998mq}.
Based on these figures, it is very plausible that our data agrees
with perturbation theory at short distances, but  data at smaller 
lattice spacing is needed for a more stringent test.

Such vacuum correlators are of course interesting in their own right.
In particular, they can be used to test  models for low-energy QCD, such
as the instanton-liquid model~\cite{Schafer:1995pz}
or the more recent holographic models of hadrons~\cite{Erlich:2005qh}.
See~\cite{Schafer:1994fd,Schafer:2007qy} for model calculations of gluonic 
vacuum correlators. A detailed comparison of the latter with lattice data  will be carried out 
elsewhere; for the time being we simply note that around $r=0.4$fm the lattice data points 
lie on a convex curve in the scalar case, and a concave curve in the pseudoscalar case. 
This observation is qualitatively consistent with the instanton-model calculations
of~\cite{Schafer:1994fd}, where it is interpreted as an attraction/ repulsion
respectively in the scalar and pseudoscalar channels.

\subsection{Finite-temperature correlators\la{sec:FTC}}
The finite-temperature correlators of the trace anomaly 
minus their zero-temperature counterpart are displayed in
\fig(\ref{fig:action2}). Partial results for this quantity
were already published in~\cite{Meyer:2008dt}. 
A sample of the raw data is presented in Tables (\ref{tab:6.2822})
and (\ref{tab:6.408}).

Starting from the higher temperatures (bottom panel), 
we note that the lattice correlators are gradually approaching the 
free-theory prediction (\eq\ref{eq:ththPT}) as the temperature is raised, as one expects
on the basis of asymptotic freedom. In order to display \eq(\ref{eq:ththPT}) on 
the figure, we have set
the renormalization scale to $\mu=\pi(T+\frac{1}{r})$ and used 
$\Lambda^{(N_{\rm f}=0)}_{\overline{\rm MS}}r_0=0.602(48)$~\cite{Capitani:1998mq}.
However, at the temperatures where simulation data are available, 
the subtracted correlator is negative at all separations $r$ 
-- unlike the free correlator.
This means that the plasma screens the fluctuations 
of the operator $\theta$ more than the vacuum does.

We have checked for discretization errors by calculating 
$G_{\theta\theta}(r,1.24T_c)$ at several lattice spacings.
This means that $\Nt=(aT)^{-1}$ is varied and $\beta$ tuned 
so that the temperature is kept fixed.
The result is shown in \fig(\ref{fig:edotb-1p24}), top panel.
To a good approximation, the data fall on a single curve.
It is not presently our intention to carry out a systematic
continuum extrapolation of $G_{\theta\theta}(r,T)$ in $1/\Nt^2$.
Rather we want to show convincing evidence that 
the conclusions we draw from finite lattice spacing data 
are not affected by discretization errors.

In the figures, we have not included the statistical uncertainty 
on the non-perturbative normalization factors. 
However the latter cancel in the ratio
\be
\frac{G_{\theta\theta}(r,T)}{ (e-3p)^2}\,,
\ee
which is displayed in \fig\ref{fig:Fsq}, top panel. 
We have multiplied this quantity by $d_A$ 
so as to give it a finite limit when $N_c\to\infty$
and by $(Tr)^4$, so that the expected short-distance behavior
is $\sim\alpha_s^2(\mu)$, where $\mu={\rm O}(1/r)$.
The graph shows that $G_{\theta\theta}(r,T)$ falls off 
like $1/r^4$, as expected from the OPE, for
$3<2\pi Tr<5$. Beyond this interval, it falls off faster to zero.

The on-shell correlation functions of $\theta$ are renormalization-group
invariant. However, if we want to compare the result with the $\N=4$ SYM 
correlator of $F^2$ described in section (\ref{sec:sym}), 
we have to divide out the factor of $\beta(g)/2g$ that multiplies 
the Yang-Mills operator $\theta$. Since the beta-function is scheme-dependent,
this means that the renormalized $F^2$ correlator in the Yang-Mills theory
is scheme-dependent, too. We choose the three-loop $\overline{\rm MS}$ 
scheme, and evaluate the coupling $\alpha_s$ at the scale 
$\mu(r,T)=\frac{3}{8}\pi (T+1/r)$. A few remarks may be useful to 
motivate this choice. At $r=1/T$, this corresponds to 
$\mu=\frac{3\pi}{4r}$; it
makes the one-loop prediction for the $\theta$ correlator approximately 
go through the lattice data at $T=0$, see \fig(\ref{fig:edotb-h}). 
At short distances, we expect $r$ to provide the harder scale
and therefore the appropriate $\mu$ to be dominated by $r$.
For $r>T$, we expect the momenta of the two exchanged gluons
to be of order $\pi T$. The chosen expression for $\mu$ 
is then a simple interpolation between these two regimes.

We thus obtain the lattice data for the $F^2$ correlator 
in the $\overline{\rm MS}$ scheme, see the bottom panel of \fig\ref{fig:Fsq}.
We are then in a position to carry out a parameter-free comparison with the 
one-loop result~\eq(\ref{eq:ththPT}) and the AdS/CFT result.
In the range of temperatures $1.2<T/T_c<1.9$, 
the lattice correlators are in semi-quantitative agreement with the 
corresponding $F^2$ correlator calculated in the strongly 
coupled ${\cal N}=4$ SYM theory.
The lattice data are negative at all $1/2T<r<1/T$, and this is in
contrast with the free-field theory result, which is positive
in that range. The data at $T>2T_c$ however does suggest that the 
non-perturbative correlator gradually moves towards the 
one-loop result as the temperature increases, as expected 
from asymptotic freedom.

Coming back to \fig\ref{fig:action2} (top panel),
the thermal fluctuations become stronger
as the temperature is lowered toward $T_c$, 
to the point where the subtracted correlator is positive
over a wide range of distances $r$.
Unsurprisingly this fact is accounted for neither by the 
weak coupling predictions, 
nor by the conformal ${\cal N}=4$ SYM result.
The correlation between the fluctuations of 
$\theta$ is strongest at $1.01T_c$, and drops
again as one moves away from the transition below $T_c$.
The interpretation of the data is helped by the fact that 
the overall fluctuations of $\theta$ are 
related to thermodynamic properties via the 
sum rule~\cite{Ellis:1998kj,Meyer:2007fc}
\be
\int d^4x \,\<\theta(x) \theta(0)\>_{{\rm conn},T-0}
= T^5 \frac{\partial}{\partial T}\frac{e-3p}{T^4}.
\la{eq:sr}
\ee
Since we know that $(e-3p)/T^4$ rises very steeply
between $T_c$ and $1.1T_c$~\cite{Boyd:1996bx}, \eq(\ref{eq:sr})
indicates that the fluctuations of $\int d^4x\,\theta$ are strongest
in that range of temperatures. 
Since the Wilson coefficients in the OPE are negative, 
there is a negative contribution to the LHS of \eq(\ref{eq:sr})
from the short-distance part of the correlator.
Therefore there has to be an enhancement in the $\theta$ two-point function
at intermediate or long distances in order to 
account for the positive sign of the RHS
(note however that here we restrict ourselves to equal-time correlators).
On the other hand, above $1.13T_c$, where the RHS of  \eq(\ref{eq:sr})
is negative, there is no necessity for $G_{\theta\theta}$
to be positive at any non-vanishing separation.
Our data shows that indeed $G_{\theta\theta}$ at intermediate distances 
$r\approx 1/T$ is negative for all available temperatures above $1.2T_c$.
This differs from the correlator $G_{ee}$ 
of the energy density~\cite{Meyer:2008dt},
which  is positive at intermediate separations.



\subsubsection{Topological charge density correlator\la{sec:TCD}}
We now discuss the topological charge density correlator, starting 
from the high temperature end.
An example of $G_{qq}$ at high-temperatures is displayed in
\fig(\ref{fig:edot-6p408}). We see that qualitatively, 
the correlator, in the interval where data is available,
resembles its free-theory counterpart:
$-G_{qq}$ is negative at short-distance, as predicted 
by the OPE, and then positive at intermediate distances.
On the basis of the pseudoscalar screening masses,
we expect $G_{qq}$ to approach zero from below, 
and there is a hint at $3.30T_c$ that indeed it does.

If we now lower the temperature, as shown in \fig(\ref{fig:qq2}, bottom panel),
we see that the range of distances where $-G_{qq}$ is positive
grows, and that its maximum value also grows. 
The OPE tells us that at sufficiently short distances, 
$-G_{qq}$ must be negative. Below $1.6T_c$ its maximum  is no longer 
visible in the data; we conjecture that it is located at too short
distance $r$ for us to see it in the lattice data
(at short distances, we are limited by discretization errors).

As we lower the temperature further (top panel of \fig(\ref{fig:qq2})),
$-G_{qq}(r,T)$ at fixed $r$ continues to grow.
It hits a maximum between $1.02$ and $1.06T_c$.
Thus similarly to the trace anomaly, the  topological charge density
exhibits strong spatial correlations near $T_c$.
How strong they are is better illustrated by taking the ratio of the 
finite-temperature to the zero-temperature data, 
\fig(\ref{fig:edotb-ratio}).
Here one clearly sees that for $r$ of order $1/T$, the spatial 
correlation of topological charge density fluctuations is about twice 
as strong near $T_c$ as in the vacuum. A technical advantage
of this ratio is that the overall normalization cancels out, and 
secondly that we expect a partial cancellation 
of the discretization errors to take place.


To check that these large correlations are not a cutoff effect --
there is after all a large cancellation taking place
at short distance in the subtracted correlator --
we have repeated the $1.24T_c$ calculation at 
a finer lattice spacing. The comparison is shown
in \fig(\ref{fig:edotb-1p24}). We see that the different data sets
fall on top of eachother within errors in the interval
$0.4<Tr<1.2$. We thus conclude that the strong, finite-temperature
induced enhancement of the correlation is a true physical effect.


There are significant differences between the scalar 
and the pseudoscalar channels in the lattice data at 
distances $1/2T<r< T$.
This is unlike the strongly coupled ${\cal N}=4$ SYM theory,
where the $F^2$ and $F\tilde F$ correlators are identical.
It is also unlike the free field prediction.  In the 
OPE framework, this difference requires operators of dimension 6 or higher
to overwhelm the expectation value of the stress-energy tensor
in \eq(\ref{eq:OPEth})\footnote{Another possibility is that 
the Wilson coefficient of $\theta$ on the RHS of the OPE 
changes sign when $r$ is not asymptotically small. This 
would signal a breakdown of the perturbative series for the 
Wilson coefficients.}. This would in turn imply the breakdown 
of the OPE as an asymptotic expansion.

Our discovery of large spatial correlations 
in the topological charge density fluctuations
in the vicinity of $T_c$ is qualitatively in agreement 
with the results of~\cite{Lucini:2004yh}.
The latter showed a strong suppression of the topological
susceptibility just above  $T_c$, using a method based on the 
semi-classical identification of topological charges.
This suppression is particularly dramatic at larger $N_c$ values, 
but even for SU(3) it amounts to  a factor of about  0.54(4).
This implies that $-\int d^4x \<q(x)q(0)\>_{T-0}\geq 0$,
and therefore there has to be a range of separations $x$ 
where $-\<q(x)q(0)\>_{T-0}$ is positive; this is what we are seeing
in the data. Note that the short-distance singularity 
of $\<q(x)q(0)\>_{T-0}\propto\alpha_s^2/x^4$ gives a finite contribution
when integrated over space-time. This is in contrast 
with the topological susceptibility itself,
$\chi_t\equiv \int d^4x \<q(x)q(0)\>_{0}$, 
which has to be defined with care~\cite{Luscher:2004fu}
if it is to remain finite when the cutoff is removed.



\section{The effective coupling in the plasma\la{sec:coupling}}

We have found that the strongly coupled SYM theory 
has an $F^2$ correlator similar to the pure Yang-Mills theory
in the deconfined phase below $2T_c$. 
To summarize the procedure, we have calculated the $\theta$ correlator on the lattice, 
which contains a factor $(\beta(g)/2g)^2$ relative to the $F^2$ correlator. 
This factor makes it renormalization-group invariant
in the pure Yang-Mills theory.
We used the 3-loop $\overline{\rm MS}$ scheme 
for the beta-function to convert the lattice $\theta$ correlator
to the $F^2$ correlator, and  found semi-quantitative agreement 
between the theories in a range of temperatures.
For $r=1/T$, the values for our chosen running coupling are
\be
\alpha_s(T)= 0.33,~~ 0.30,~~ 0.27,~~ 0.25,~~ 0.23,~~ 0.19
\la{eq:alphas}
\ee
for the six temperatures
displayed on the bottom panel of \fig(\ref{fig:action2}).
From \fig(\ref{fig:Fsq}), we see that at the last two temperatures,
the Yang-Mills $F^2$ correlator no longer agrees with the strongly 
coupled SYM result. Therefore we conclude that for $\alpha_s$
smaller than about $0.25$ (i.e. $\lambda$ smaller than about 10), 
one should not expect other properties of the Yang-Mills plasma 
to coincide with those of the infinite-coupling SYM plasma.
In fact it is somewhat surprising that 
for $0.25<\alpha_s<0.30$, where the function $\alpha_{\overline{\rm MS}}(\mu)$
shows a modest dependence on the order in perturbation theory,
the thermal correlator is so much more similar to the strongly coupled 
SYM correlator than to the weakly coupled one.
This may be related to the fact that at finite temperature, 
due to infrared effects, the perturbative expansion parameter
is $g$ rather than $\alpha_s$.
Thus at an energy scale where the vacuum polarization effects are 
still well approximated by the perturbative expansion, 
the thermal physics rather has a strong coupling character.

We now discuss whether one may 
use  the `empirically' observed similarity 
to match the couplings of the Yang-Mills and SYM theories.
By `matching', we  mean to find a way of
comparing the two theories in such a way that they share
as many properties as possible at a semi-quantitative 
level~\footnote{This notion is similar to, but distinct from the 
(more precisely defined) matching procedure used in effective field theories.}.
Since  the 't~Hooft coupling is the unique parameter of the 
SYM theory, this is the only parameter we need to fix 
in the comparison.
To what extent several observables can be simultaneously 
made similar provides a clue as to how universal the 
properties of non-Abelian plasmas are.

The best way to match QCD with a different theory is presumably 
to equate a renormalized quantity such as the Debye mass~\cite{Bak:2007fk}
across the two theories. 
However this requires knowing the relation between the coupling and the Debye
mass on the SYM side. A technical obstacle to this program is
that the Debye mass is independent of the coupling
in the limit of large coupling. Other observables
typically lead to the same lack of sensitivity to $\lambda$.
One then needs to know $1/\lambda$ corrections to the
selected observable on the SYM side, which leads to
more involved calculations and raises questions
of convergence, etc.

A different way to match the two theories is to
define a running coupling based on a renormalized quantity, 
such that the weak-coupling relation holds by definition 
for all scales.
One then equates the couplings of the two theories.
For example, one can define a Yang-Mills effective coupling from the 
Debye mass, $\lambda(T)\equiv 3\frac{m_D^2}{T^2}$
in the SU($N_c$) gauge theory, and use that value of $\lambda$
in the SYM theory.  A priori, when the coupling
is large, its scheme dependence is strong.
For this reason, we expect that matching the observable itself 
is the superior procedure.

Nevertheless the non-trivial agreement of the Yang-Mills
and the SYM $F^2$-correlators in a range of temperatures suggests 
that simply using the values of $\lambda=12\pi\alpha_s=10\dots12$,
where the temperature-dependent 
values of $\alpha_s$ are given in \eq(\ref{eq:alphas}),
is a reasonable choice of coupling constant to use on the SYM side.
This is a moderately large coupling constant; for instance, 
the  O($1/\lambda^{3/2}$) correction to the $\lambda=\infty$
shear viscosity to entropy density ratio $\eta/s$ 
is about $+50\%$ at this coupling~\cite{Buchel:2004di,Buchel:2008sh,Myers:2008yi}.

\section{Summary\la{sec:disc}}

We have found that the gluon plasma generically 
screens scalar and pseudoscalar
fluctuations more than the vacuum does at  
short distance $r\ll 1/T$ and at long distances $r\gg 1/T$.
Near $T_c$ however, there is a significant range of distances
of order $1/T$  over
which the spatial correlations are stronger in the plasma
than in the vacuum. We interpret this fact as there being
stronger fluctuations of wavelength O($1/T$) in the plasma
than in the vacuum. As one increases the temperature above 
$T_c$, this effect disappears soon in the scalar channel, 
but extends to about $2T_c$ in the pseudoscalar case.
In the latter channel, the enhancement of these fluctuations
over those of the vacuum is about a factor two.
While the pseudoscalar and scalar channels are expected to
have similar correlation functions at very short distances
and they precisely agree in the SYM theory, 
the two channels look rather different
at least up to $2T_c$, according to our lattice data.
The scalar correlator agrees well with the corresponding correlator 
in the strongly coupled SYM theory in the range of temperatures $1.2<T/T_c<1.9$,
while the pseudoscalar correlator is notably different due to
the aforementioned strong fluctuations. These observations 
constitute our main results. The scalar fluctuations of 
wavelength $\sim 1/T$ are suppressed compared to the vacuum,
while the pseudoscalar fluctuations are significantly enhanced.
It would be interesting to see whether a next-to-leading order 
perturbative calculation would agree significantly better with 
the lattice data than the treelevel calculation does. 

We note that studying the vacuum subtracted correlators
of gauge invariant operators at distances short compared to $1/T$
is morally equivalent, via the operator-product expansion,
to investigating the thermal expectation values of 
higher-dimensional renormalized operators. Our study thus has  goals in common
with the investigation of twist-two operator expectation values~\cite{Iancu:2009zz}.

The semi-quantitative agreement of the scalar correlators 
between the pure Yang-Mills and the SYM theories, while
the pseudoscalar channel is markedly different, highlights 
the fact that different plasmas can exhibit quite similar properties 
in some channels while differing substantially in others.

In spite of having a reduced topological susceptibility~\cite{Lucini:2004yh},
the deconfined phase close to $T_c$ exhibits strong 
correlations of $\vec E\cdot \vec B$ over distances
of order $1/T$, which are stronger than in the vacuum by 
about a factor two. It would be interesting to see whether
models of QCD can account for this effect.
It would also be worth investigating how much this effect
depends on the weakness of the first-order deconfining phase transition,
and whether the effect persists at larger 
values of $N_c$, where the transition is strongly first order~\cite{Lucini:2002ku,Lucini:2005vg}.
A plausible mechanism for the observed strong spatial correlations
is that fluctuations of $\vec E\cdot \vec B$  with a 
coherence length of at least $1/T$ occur in the plasma.
Perhaps these fluctuations have been seen in~\cite{Lucini:2004yh},
where the topological lump size was found to be peaked 
at $\rho\simeq1.7/T_c$. This large size led the authors
to conclude that this peak lies outside the range of applicability
of their semiclassical methods.
It is worth thinking about possible phenomenological 
implications of these large-amplitude, long-wavelength fluctuations, 
since the charge-separation
effects of a non-zero $\vec E\cdot \vec B$ field configuration
in the context of heavy ion collisions have recently received a lot of 
attention~\cite{Kharzeev:2007jp,Kharzeev:2009mf,Voloshin:2009hr}.

\FIGURE[t]{
\vspace{0.65cm}

\centerline{\includegraphics[width=10.0 cm,angle=-90]{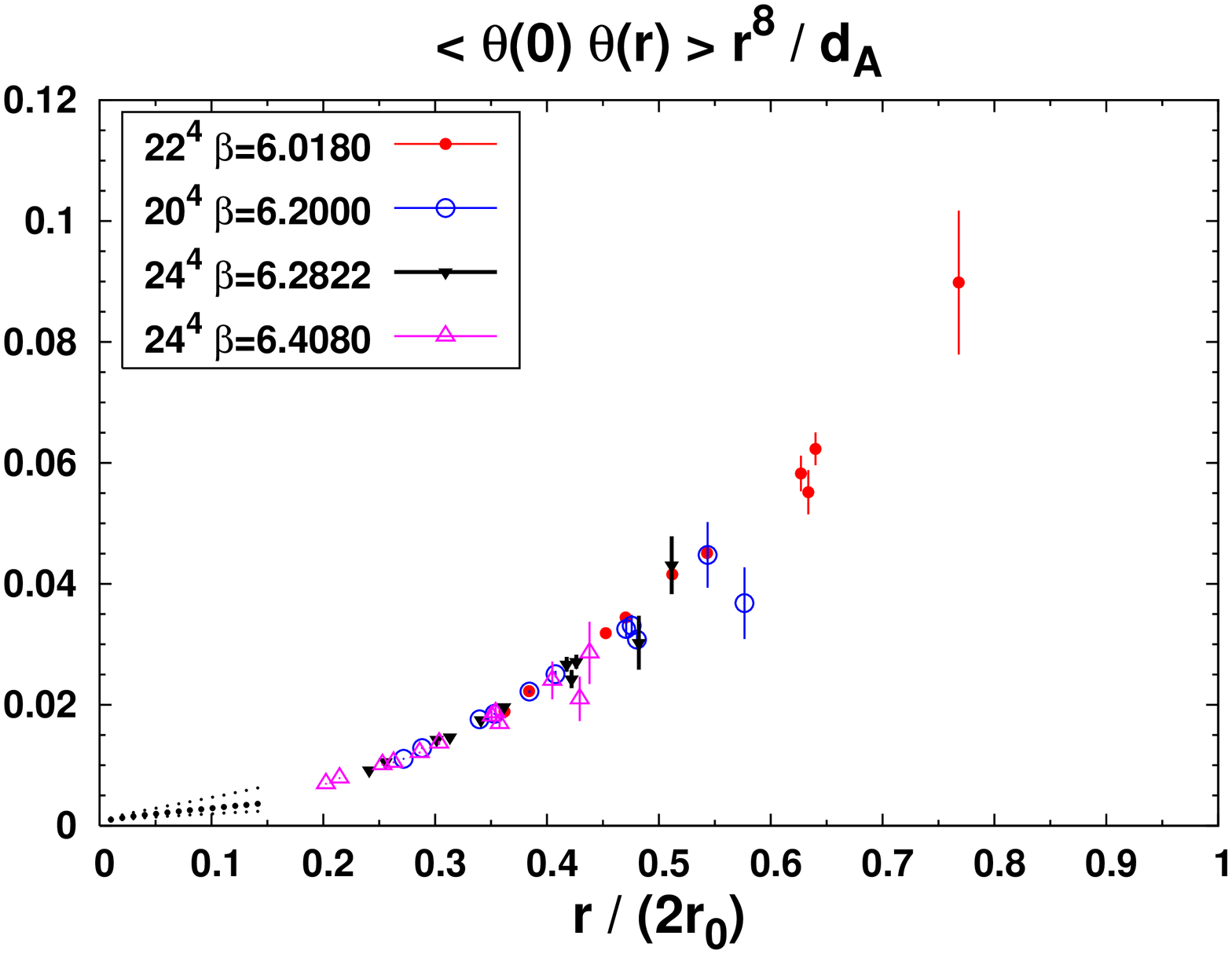}}

\centerline{\includegraphics[width=10.0 cm,angle=-90]{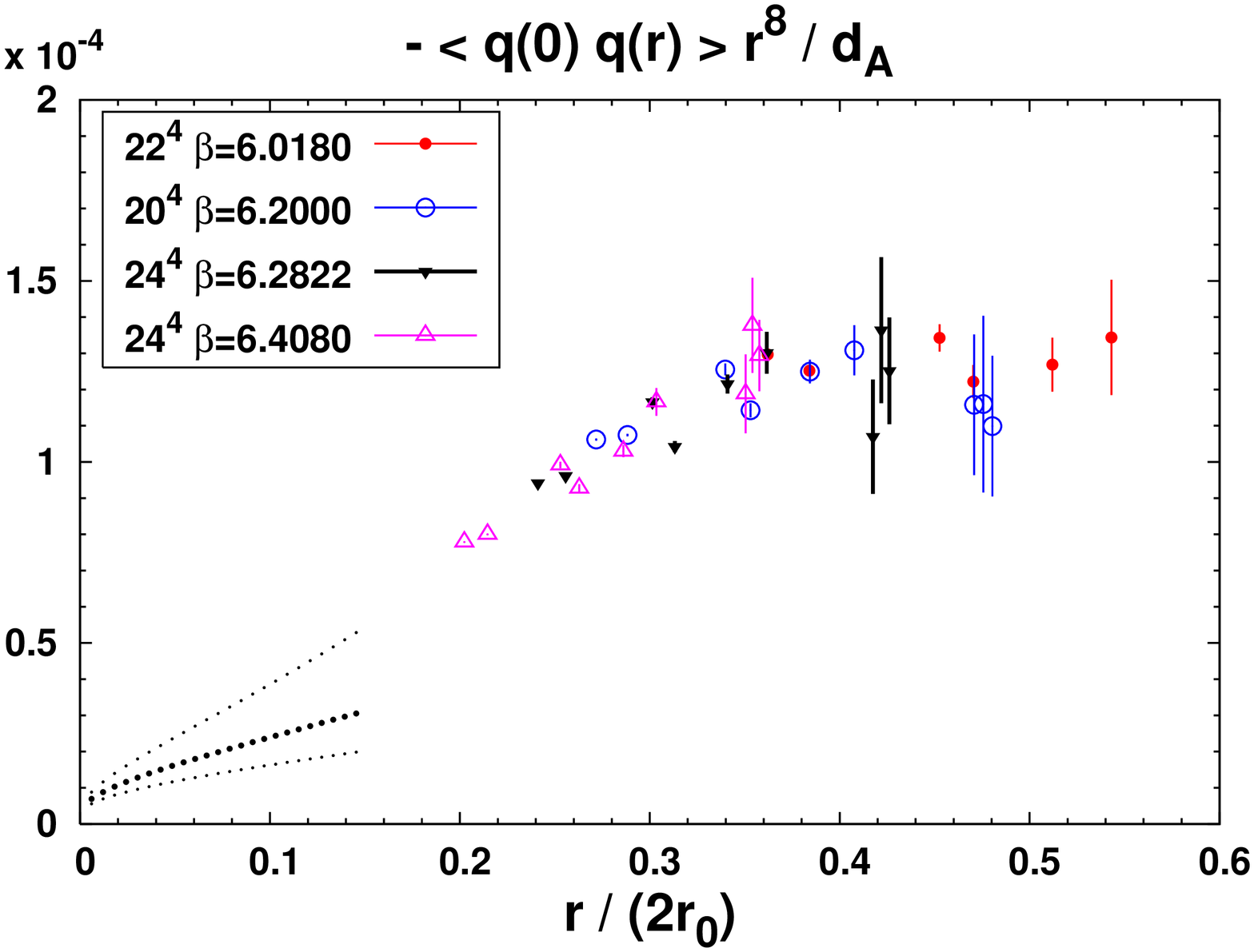}}

\caption{Zero-temperature correlator of the trace anomaly $\theta(x)$ (top)
and of the topological charge density $q(x)$ (bottom).
The overall normalization in the latter case is fixed by the data of 
Horvath et al.~\cite{Horvath:2005cv}. The dotted lines at small $r$ 
correspond to the one-loop result with choices of renormalization scale 
(from top to bottom) $\mu=\frac{\pi}{2r},\frac{\pi}{r}$ and $\frac{2\pi}{r}$.}
\label{fig:edotb-h}
}
\noindent
\FIGURE[t]{
\vspace{0.65cm}

\centerline{\includegraphics[width=8.0 cm,angle=-90]{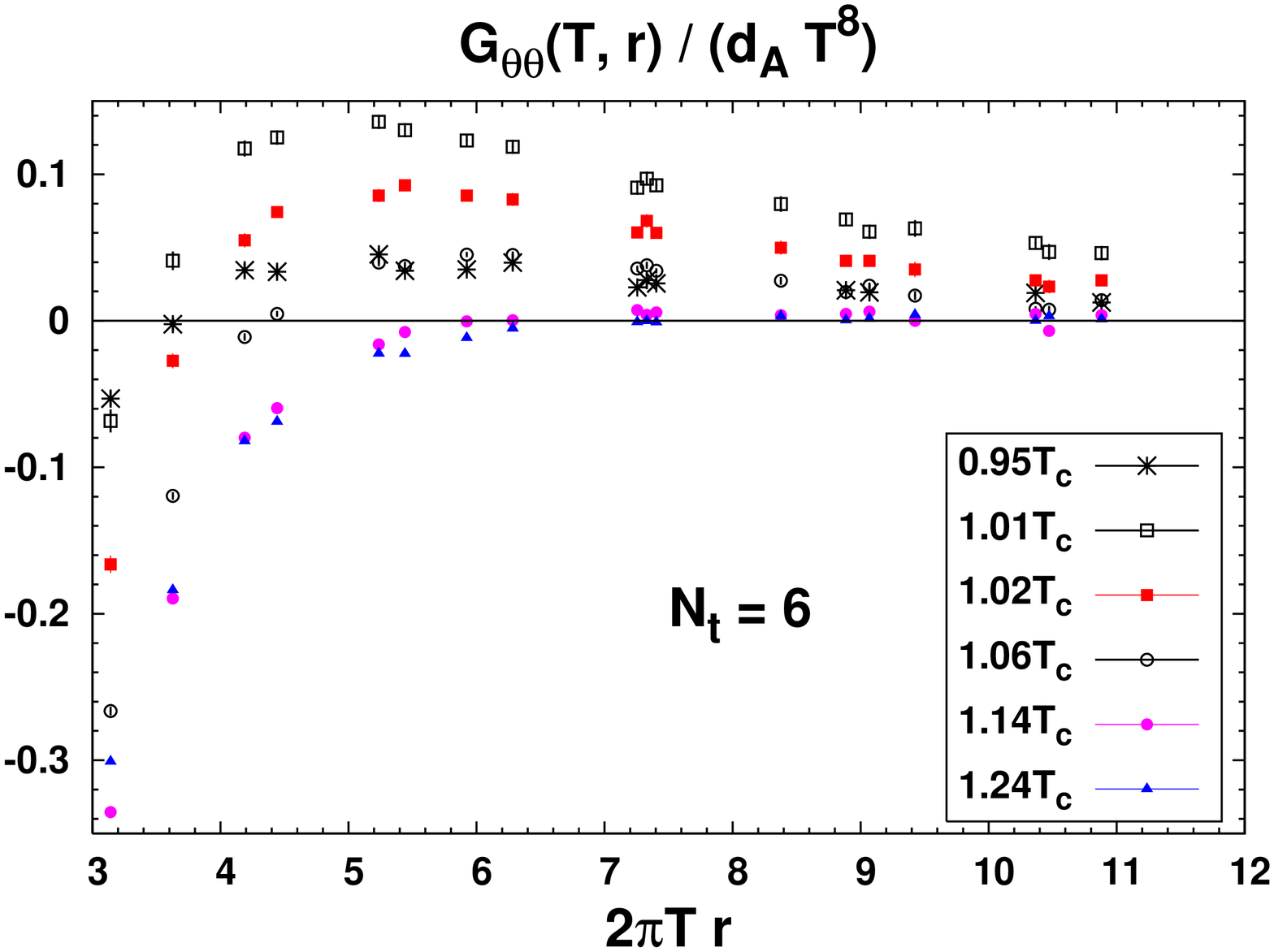}}

\centerline{\includegraphics[width=8.0 cm,angle=-90]{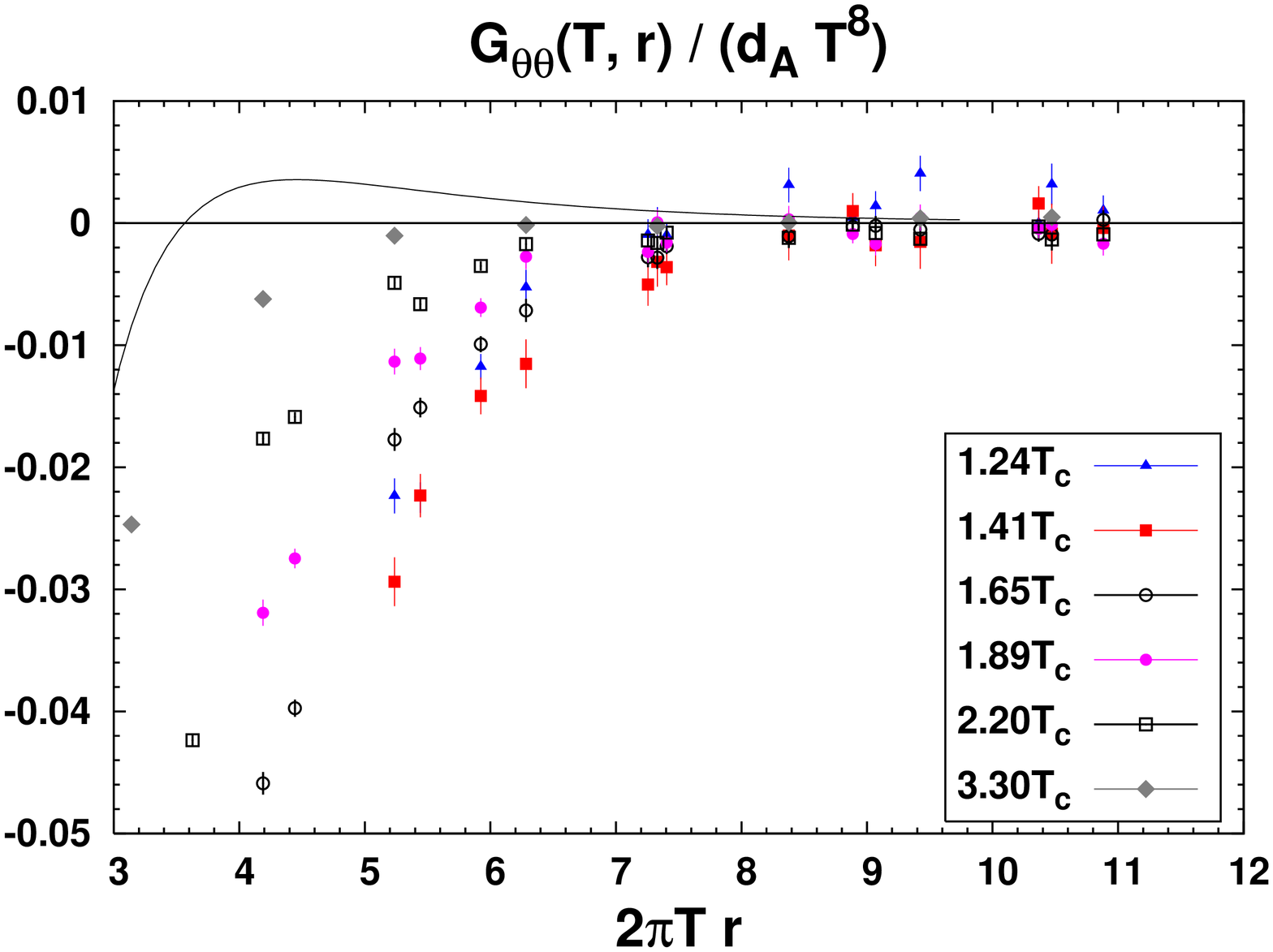}}

\caption{Thermal part of the spatial correlator of the trace anomaly $\theta$,
at $\Nt=6$ and several temperatures. 
The curve in the lower panel is the one-loop result (\eq\ref{eq:ththPT})
with choice of renormalization scale described in section (\ref{sec:FTC}).}
\label{fig:action2}
}
\noindent
\FIGURE[t]{
\vspace{0.65cm}

\centerline{\includegraphics[width=8.0 cm,angle=-90]{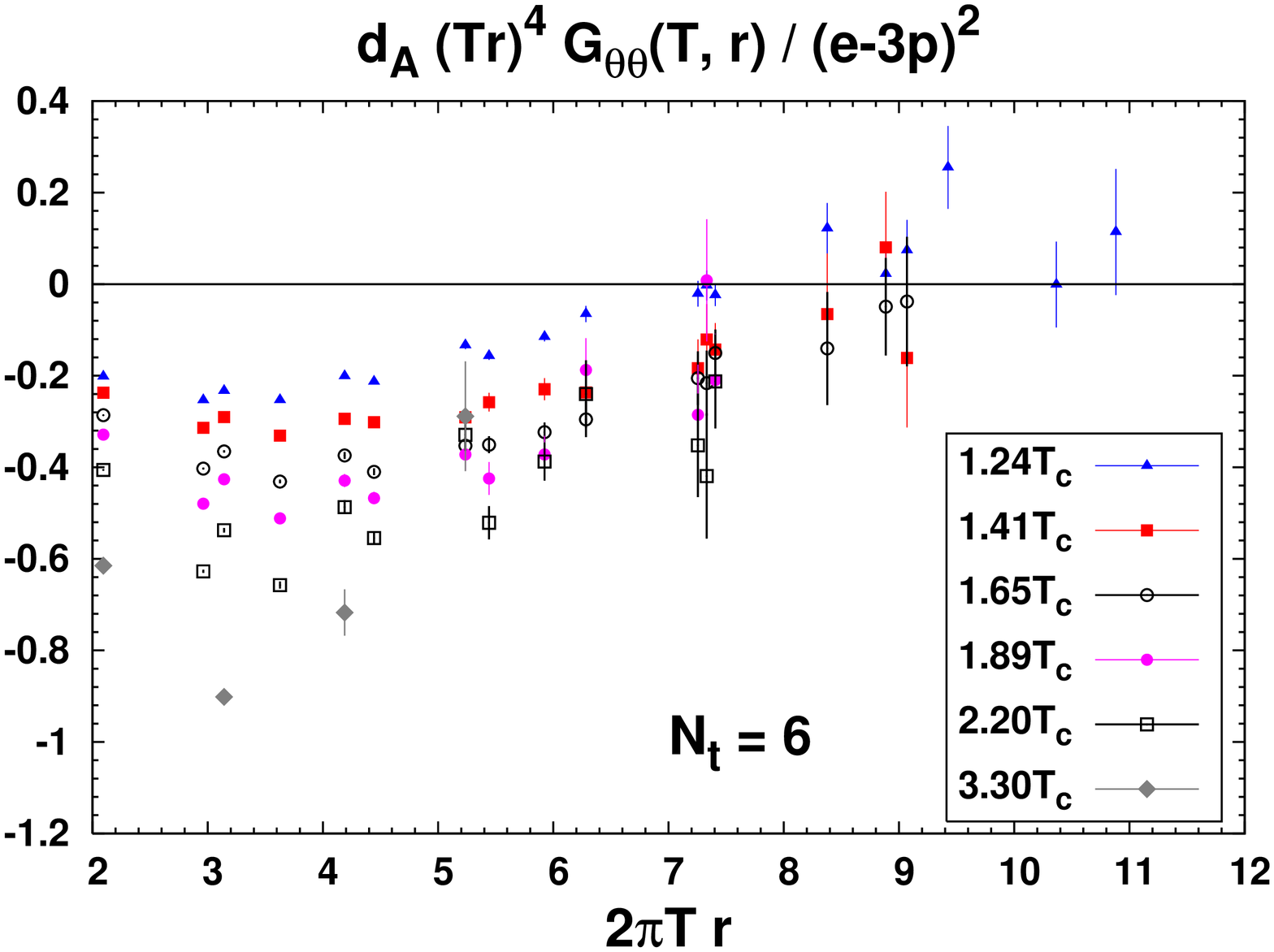}}

\centerline{\includegraphics[width=8.0 cm,angle=-90]{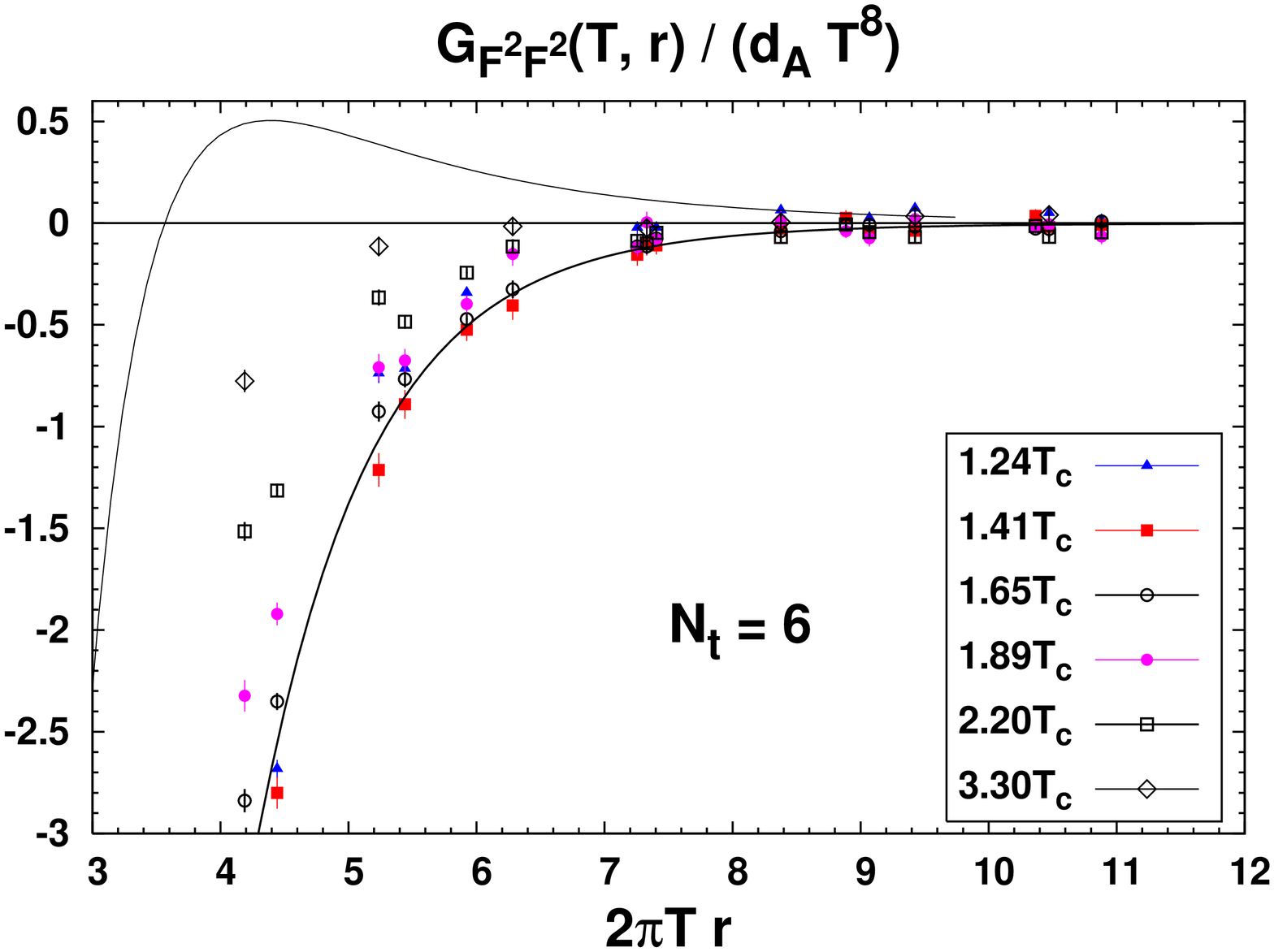}}

\caption{Top: thermal part of the spatial correlator 
of the trace anomaly $\theta$, normalized by $(e-3p)^2$
(which cancels the renormalization factor).
Bottom: comparison of the $F^2$ correlator in the 3-loop $\overline{\rm MS}$ scheme
to the one-loop result (upper curve) and the SYM correlator (lower curve).}
\label{fig:Fsq}
}
\noindent
\FIGURE[t]{
\vspace{0.65cm}

\centerline{\includegraphics[width=8.0 cm,angle=-90]{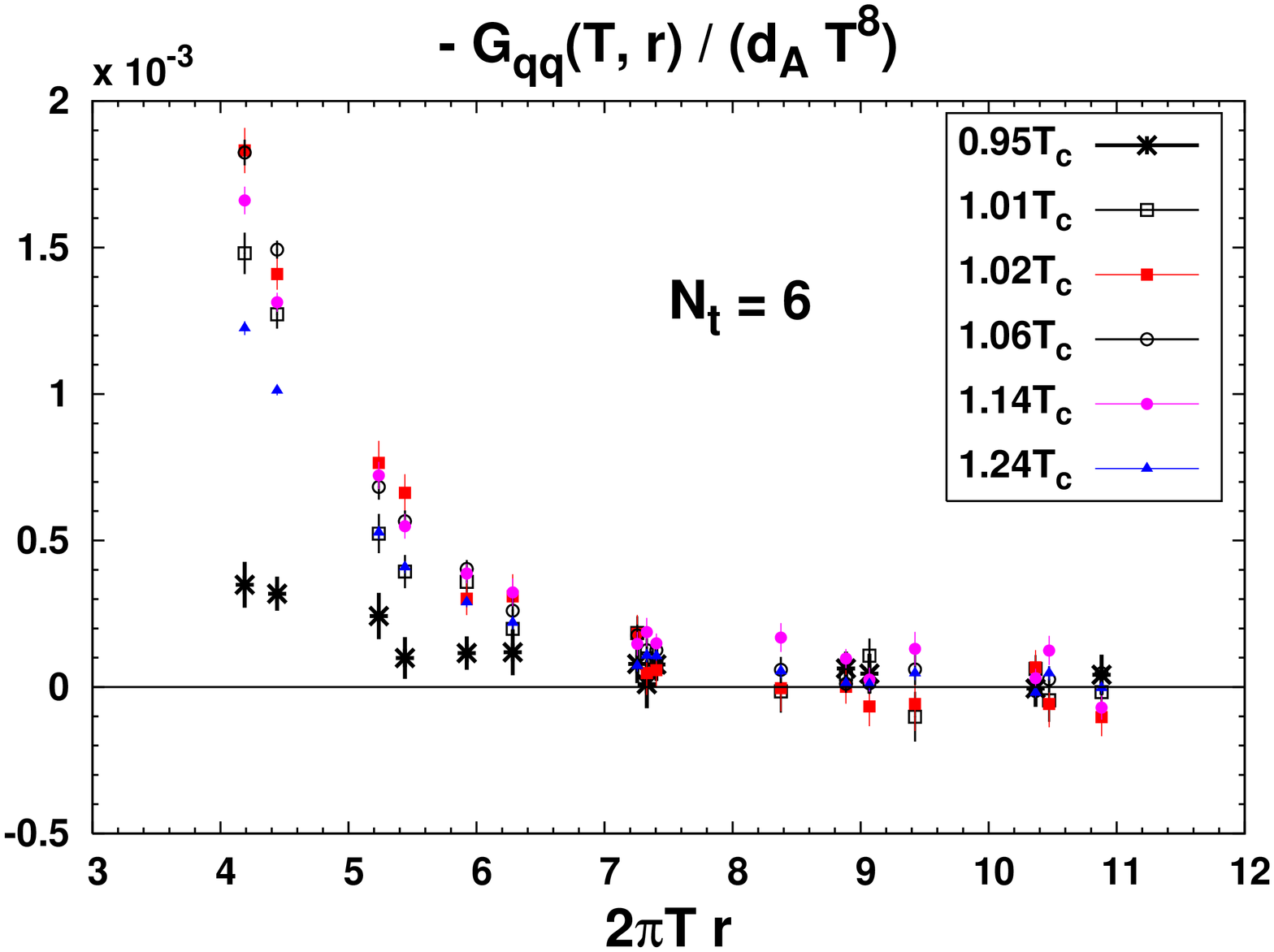}}

\centerline{\includegraphics[width=8.0 cm,angle=-90]{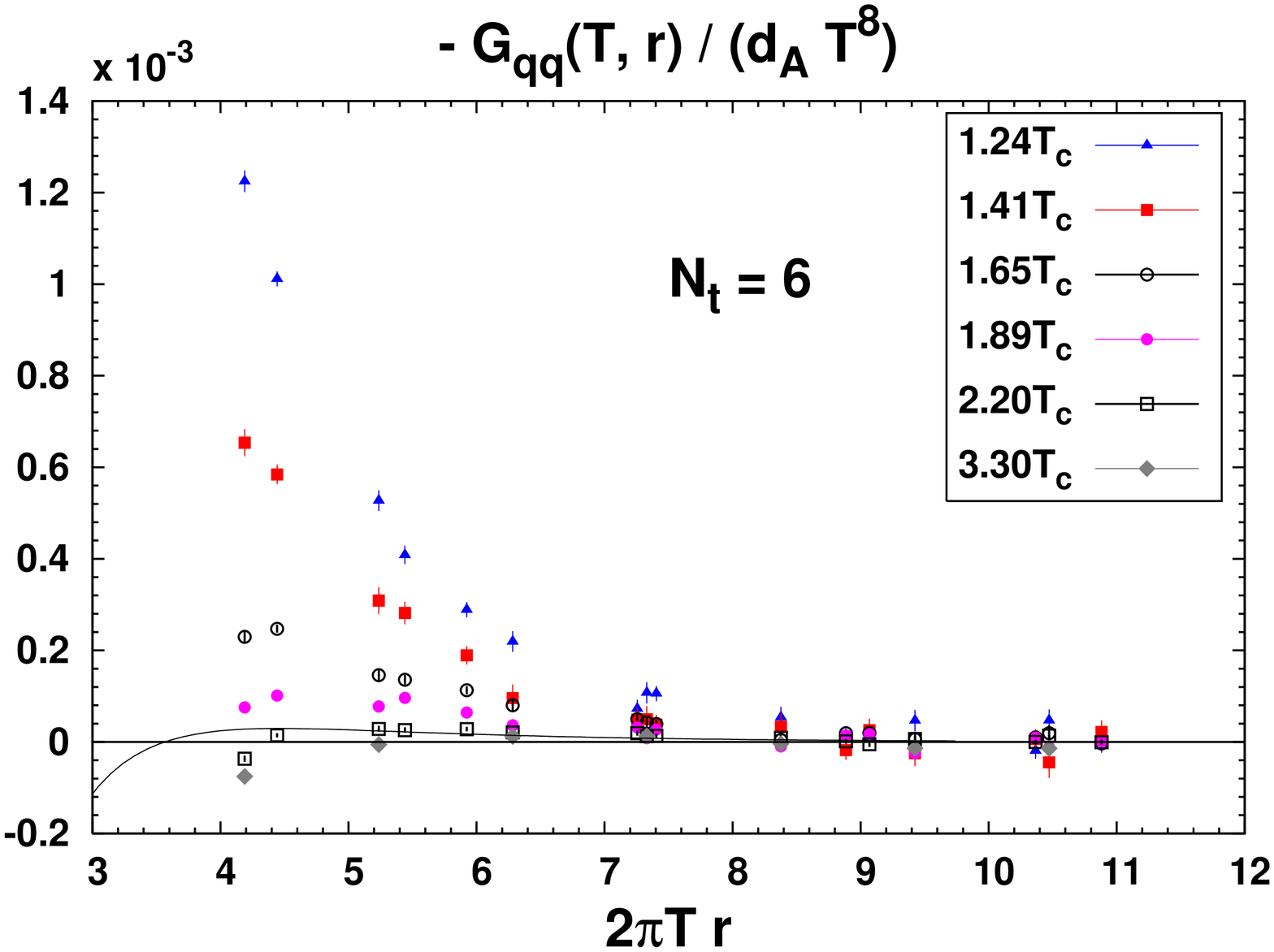}}

\caption{Thermal part of the spatial correlator of the topological charge density $q$,
at $\Nt=6$ and several temperatures. 
The curve in the lower panel is the one-loop result (\eq\ref{eq:ththPT})
with the same choice of renormalization scale as in \fig(\ref{fig:action2}).
}
\label{fig:qq2}
}
\noindent
\FIGURE[t]{
\vspace{0.65cm}

\centerline{\includegraphics[width=9.0 cm,angle=-90]{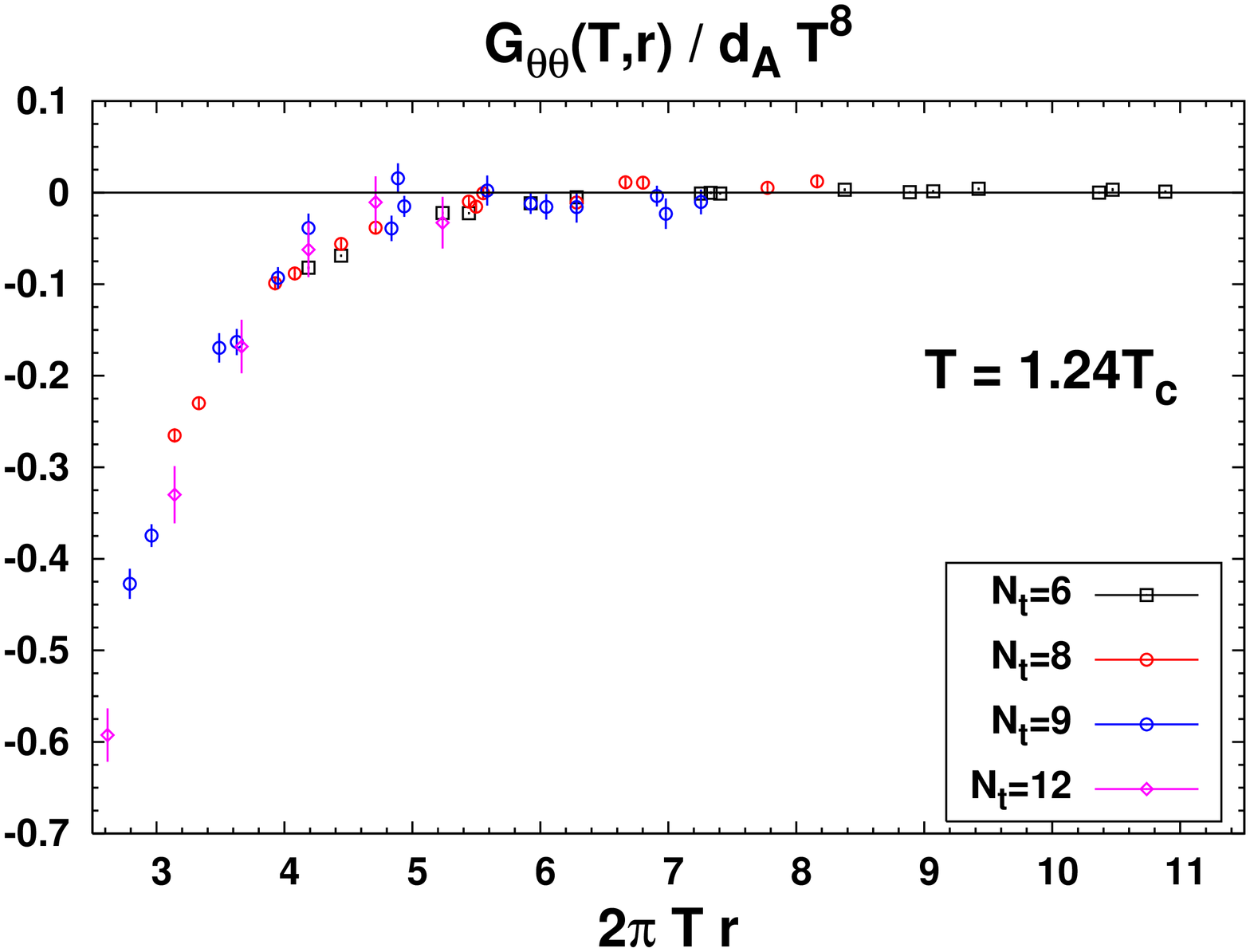}}

\centerline{\includegraphics[width=9.0 cm,angle=-90]{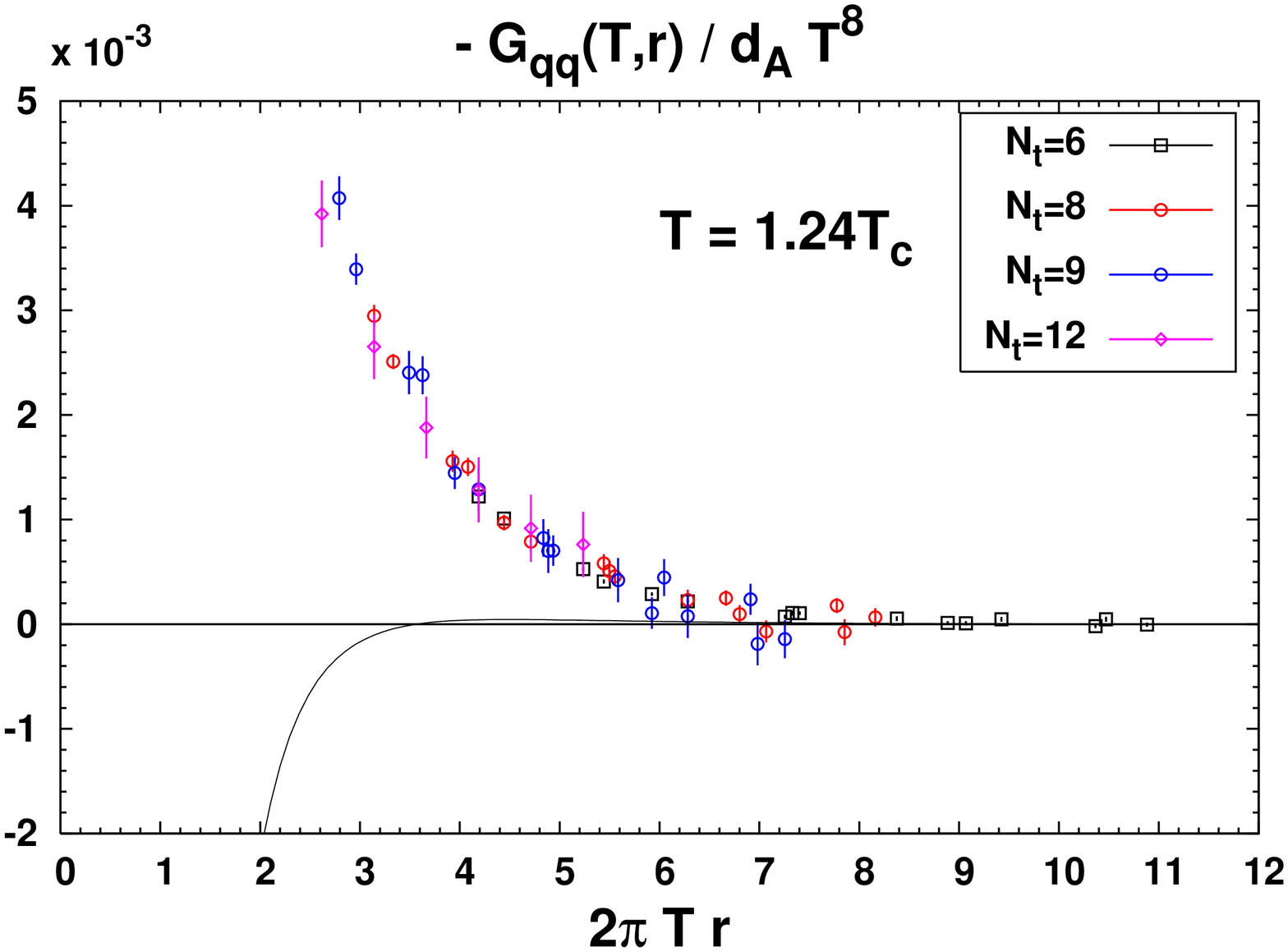}}

\caption{Thermal part of the spatial correlator 
of the trace anomaly (top) and topological charge density (bottom),
at $T=1.24T_c$. The curve in the lower panel is the one-loop result (\eq\ref{eq:ththPT})
with the same choice of renormalization scale as in \fig(\ref{fig:action2}).}
\label{fig:edotb-1p24}
}
\noindent
\FIGURE[t]{
\centerline{\includegraphics[width=9.0 cm,angle=-90]{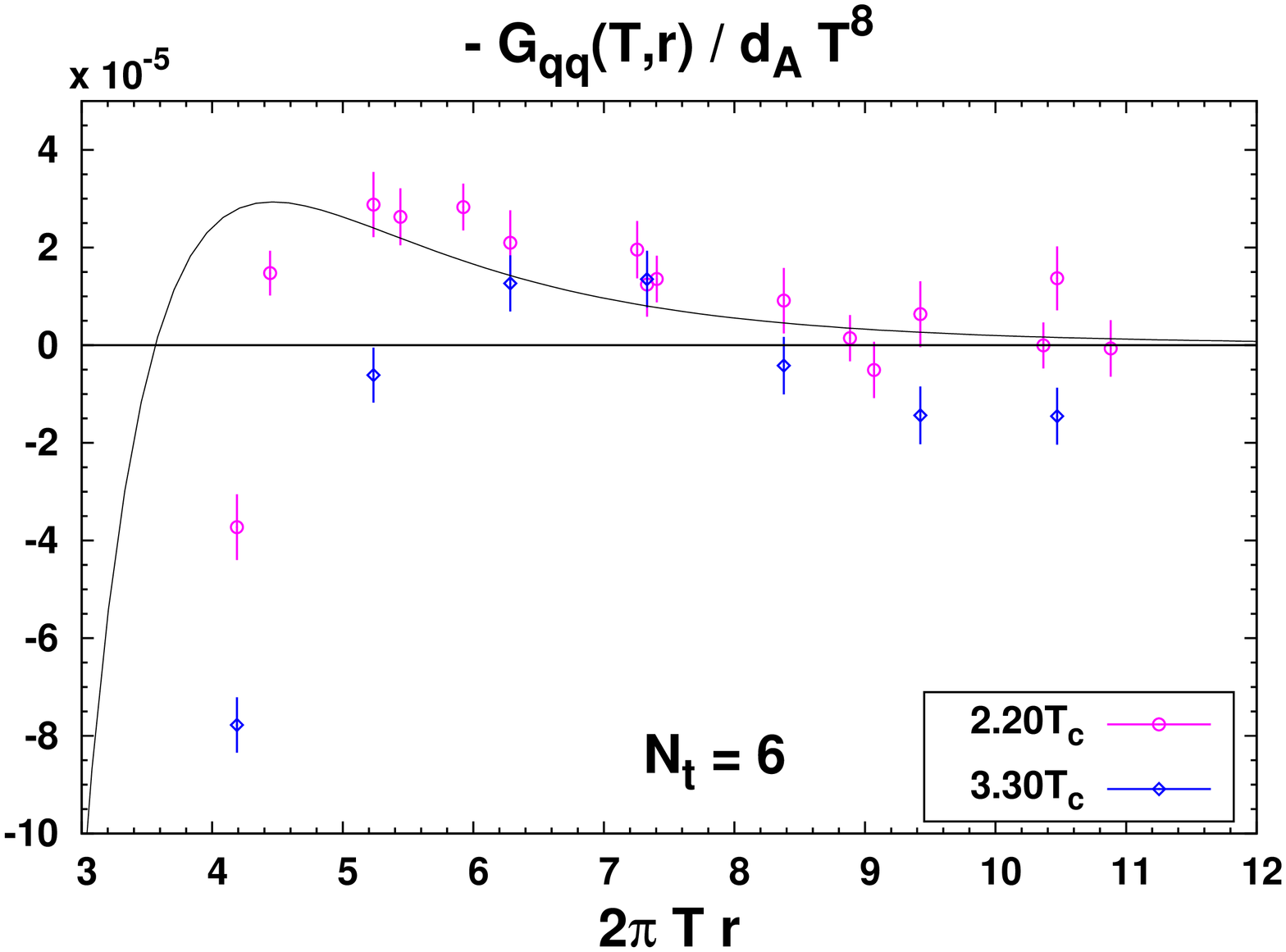}}
\caption{Detail of the thermal part of the spatial correlator 
of the topological charge density $q(x)$ at $\beta=6.408$,
compared to the one-loop result (\eq\ref{eq:ththPT})
with the same choice of renormalization scale as in \fig(\ref{fig:action2}).}
\label{fig:edot-6p408}
}
\noindent
\FIGURE[t]{
\centerline{\includegraphics[width=9.0 cm,angle=-90]{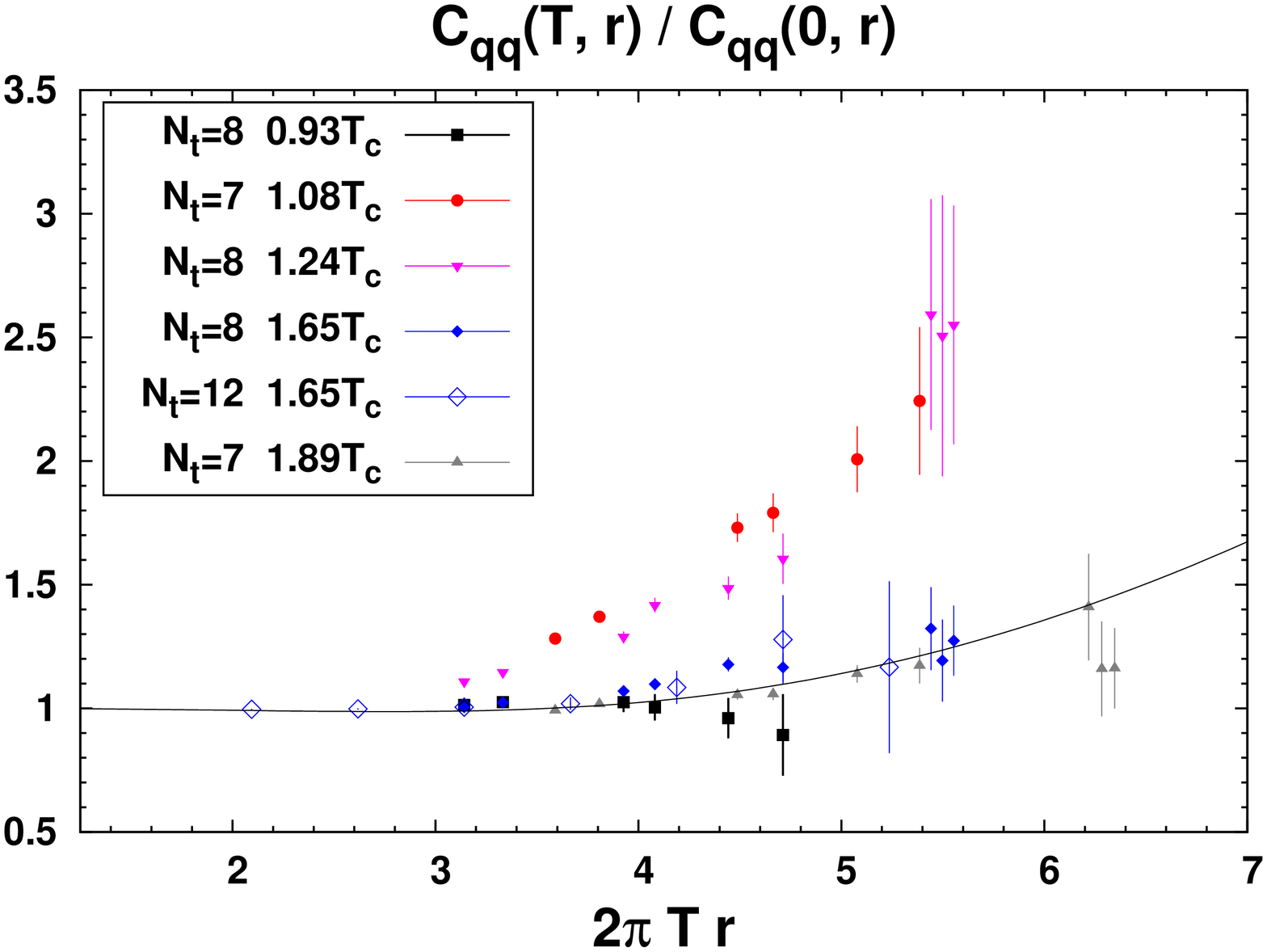}}
\caption{Ratio of  the finite-temperature spatial correlator 
of the topological charge density $q(x)$
to the zero-temperature one.
The curve is the one-loop result (\eq\ref{eq:ththPT}).}
\label{fig:edotb-ratio}
}
\noindent
\FIGURE[t]{
\centerline{\includegraphics[width=13.0 cm,angle=0]{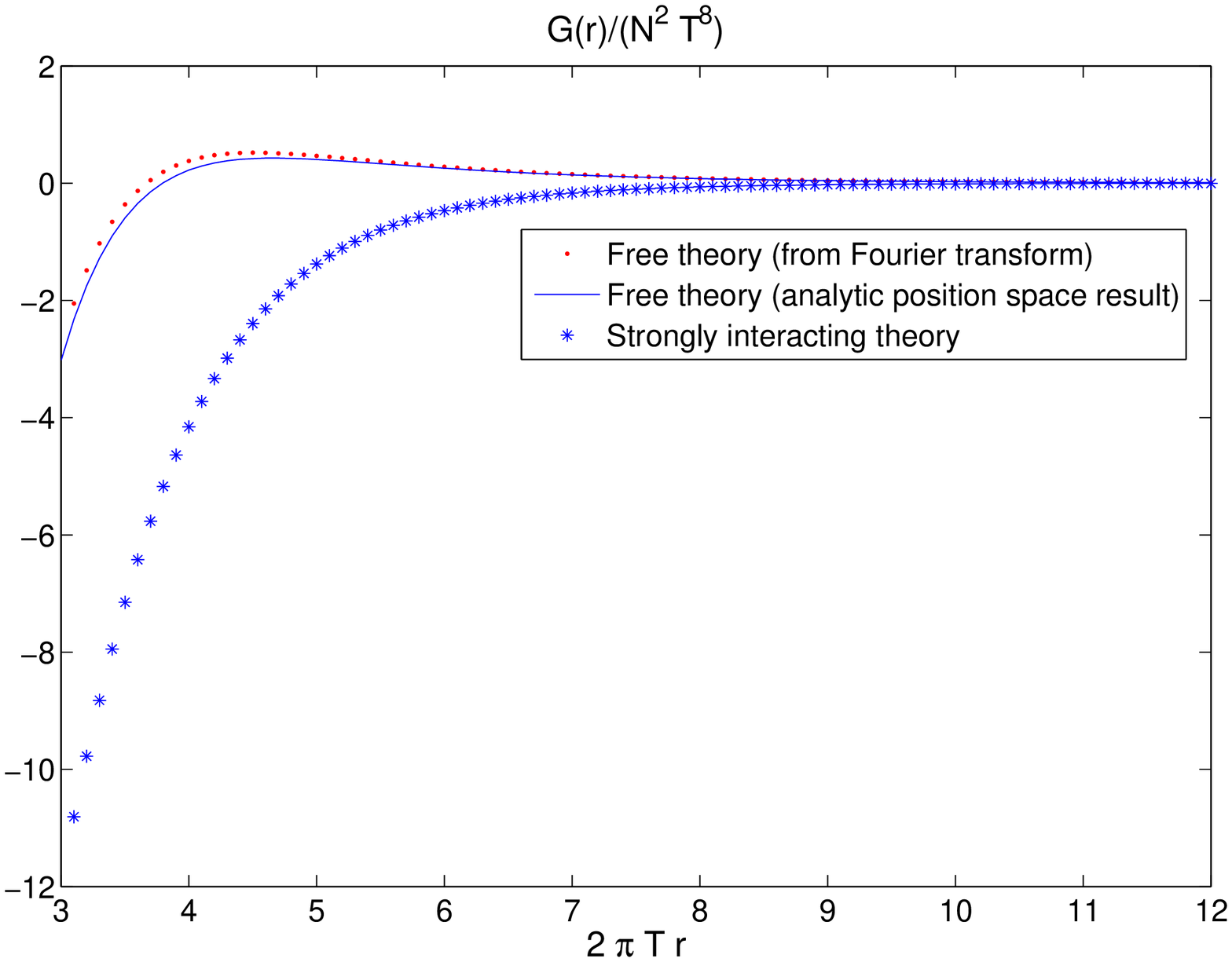}}
\caption{
SYM correlators obtained by Fourier transform of the spectral function.
}
\label{fig:justone}
}
\noindent
 \FIGURE[t]{
 \centerline{\includegraphics[width=13.0 cm,angle=0]{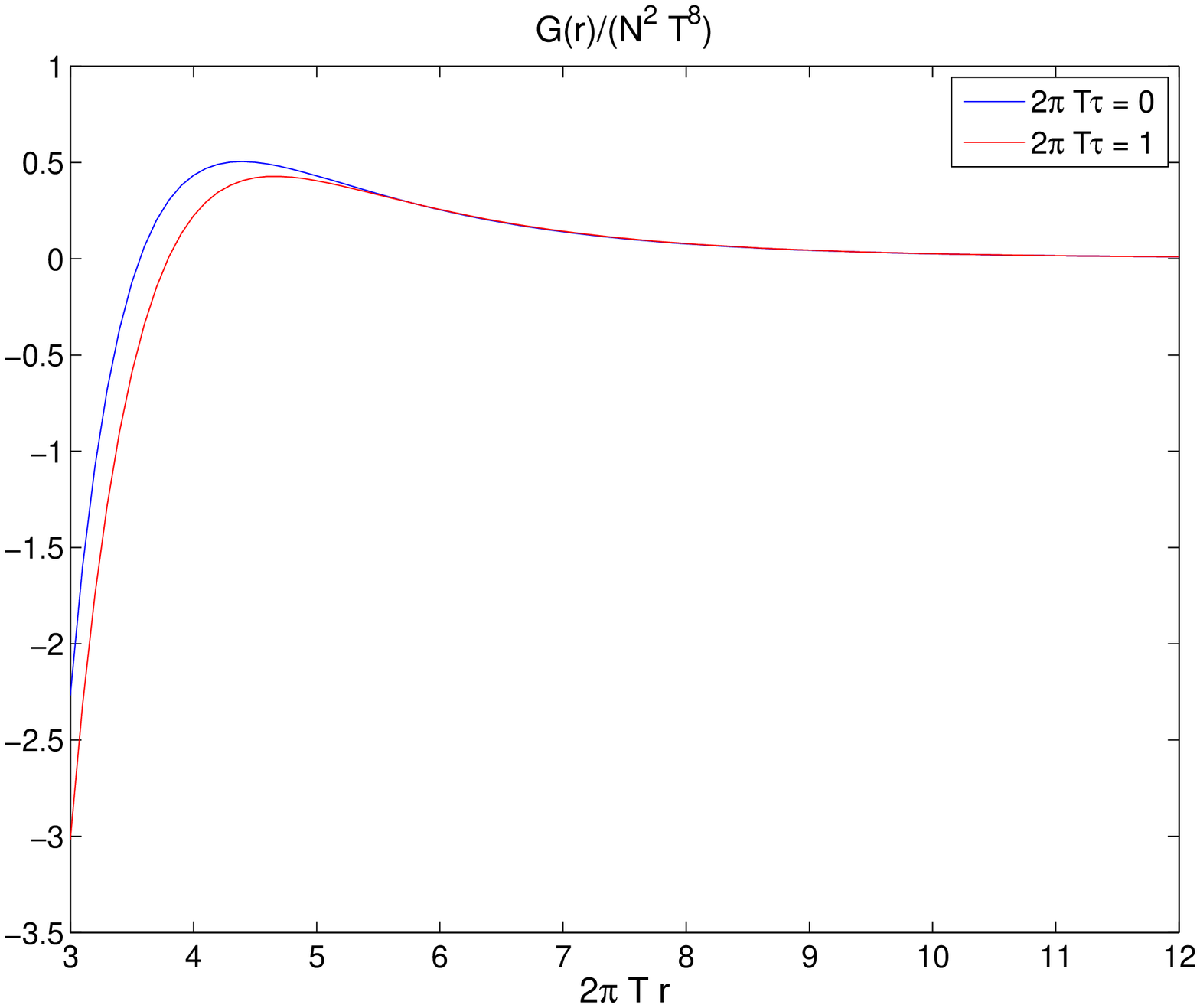}}
 \caption{
The free correlators with $2\pi T \tau= 1$ and 0, \eq(\ref{eq:free}).
 }
 \label{fig:tautest}
 }
 \noindent

\acknowledgments{
We thank K.~Rajagopal, H.~Liu, J.~Minahan and A.Vuorinen
 for useful discussions.
H.M. thanks K.F.~Liu and I.~Horvath for providing the 
raw lattice data of Ref.~\cite{Horvath:2005cv}.
Our simulations were done on a BlueGene L rack at MIT and 
the desktop machines of the Laboratory for Nuclear Science.
This work was supported in part by
funds provided by the U.S. Department of Energy 
under cooperative research agreement DE-FG02-94ER40818.
}
\appendix

\section{Details of Numerical Integration} \label{app:numerics}
We must numerically integrate the flow equation (\ref{floweqn})
\be
\partial_r\tilde{\chi} = \frac{2 i \bar{\omega}}{r^2\left(1 - \frac{1}{r^4}\right)}\left[{\tilde{\chi}^2 \ov r^3} - r^3\left[1-\frac{\bar{k}^2}{\bar{\omega}^2}\left(1-\frac{1}{r^4}\right)\right]\right]. \label{floweqn2}
\ee 
from the horizon at $r = 1$ up to the AdS boundary at $r \to \infty$, which the initial condition $\tilde{\chi}(r = 1) = 1$. In practice we integrate to $r = 20000$ and verify that further increasing the integration domain does not change the answer. Note that while $\Im(\chi)$ contains a divergence as $r \to \infty$, this is a standard UV divergence that contributes only a contact term, and can be removed by holographic renormalization. It will not concern us, as the real part of $\chi$ (and thus the imaginary part of $G_R$) has a finite limit as $r \to \infty$.

We cannot begin our integration at precisely $r = 1$ as the equation is singular there. We thus build a series expansion of $\tilde{\chi}$ about $r = 1$:
\be
\tilde{\chi} = 1 + \tilde{\chi}_1 (r-1) + \tilde{\chi_2}(r-1)^2 + ...
\ee
Plugging this into \eq(\ref{floweqn2}) we can determine the expansion coefficients $\chi_{n}$ up to any desired order. The expressions are lengthy but straightforward to obtain and so we do not present them here; however we use the first three terms in this expansion to find the value of $\tilde{\chi}(r = 1+\delta)$ and use this the initial condition to begin our integration at a small finite value of $\delta$ ($\delta = 0.01$ in practice). 

Note also that if we keep $\bar{k}$ finite and take $\bar{\omega} \to 0$, the flow equation appears singular. However we know that at vanishing chemical potential the spectral density of a bosonic operator must be an odd function of $\bar{\omega}$, and thus vanishes as $\bar{\omega} \to 0$, although the precise point $\bar{\omega} = 0$ presents numerical difficulties. Thus our code simply sets $\Im(G_R(\bar{\omega} = 0)) = 0$ by hand.


\bibliographystyle{JHEP}
\bibliography{../../../../BIBLIO/viscobib.bib}


\TABLE[!p]{
\begin{tabular}{|c|r@{~~~~}r|r@{~~~~}r|}
\cline{1-5}
  \multicolumn{1}{|c|}{ }    & \multicolumn{2}{c |}{$22^4,\,\beta=6.018$, $\frac{dg_0^{-2}}{d\log a}=-0.09922$}  
&  \multicolumn{2}{c|}{$24^4,\,\beta=6.2822$, $\frac{dg_0^{-2}}{d\log a}=-0.1166$} \\
 \multicolumn{1}{|c|}{ $r/a$}    & \multicolumn{2}{c |}{ $\chi_s=1.371$, $Z_q=1.742$}  
&  \multicolumn{2}{c|}{$\chi_s=1.220$, $Z_q=1.546$} \\
 & \multicolumn{1}{c}{ $r^8C^{\rm bare}_{\theta\theta}/d_A$}    
&   \multicolumn{1}{c|}{ $-10^4\cdot r^8 C^{\rm bare}_{qq}/d_A$}  & \multicolumn{1}{c}{ $r^8C^{\rm bare}_{\theta\theta}/d_A$} 
 &   \multicolumn{1}{c|}{ $-10^4\cdot r^8C^{\rm bare}_{qq}/d_A$ } \\
\cline{2-5}
\hline
2.0000  & 0.00156574(33) & -0.043481(18) & 0.00085488(12) & -0.0224117(68) \\ 
3.0000  & 0.0074804(48) & 0.13816(24) & 0.0037726(20) & 0.12537(10) \\ 
4.0000  & 0.018866(42) & 0.4276(22) & 0.009190(18) & 0.39384(93) \\ 
5.0000  & 0.03185(24) & 0.443(12) & 0.01424(10) & 0.4876(55) \\ 
6.0000  & 0.0451(10) & 0.443(52) & 0.01965(43) & 0.544(24) \\ 
7.0000  & 0.0552(37) & 0.46(19) & 0.0242(15) & 0.571(84) \\ 
8.0000  & 0.056(10) & -0.17(55) & 0.0303(44) & 0.56(24) \\ 
9.0000  & 0.024(27) & -3.2(14) & 0.038(11) & 1.13(61) \\ 
10.0000 & -0.045(72) & -9.0(35) & 0.042(27) & 3.5(14) \\
\hline 
2.8284  & 0.0070641(29) & 0.13736(13) & 0.0038655(11) & 0.124059(49) \\ 
4.2426  & 0.022254(50) & 0.4127(25) & 0.010561(22) & 0.4021(11) \\ 
5.6569  & 0.04157(46) & 0.418(25) & 0.01751(20) & 0.508(11) \\ 
7.0711  & 0.0623(27) & 0.12(15) & 0.0271(12) & 0.523(62) \\ 
8.4853  & 0.090(12) & 0.43(61) & 0.0431(48) & 0.64(28) \\ 
9.8995  & 0.122(41) & 3.6(22) & 0.044(17) & -0.36(94) \\ 
\hline
3.4641  & 0.013541(12) & 0.33075(62) & 0.0071350(55) & 0.29340(29) \\ 
5.1962  & 0.03443(28) & 0.403(15) & 0.01465(12) & 0.4360(66) \\ 
6.9282  & 0.0582(29) & 0.40(15) & 0.0267(12) & 0.447(66) \\ 
8.6602  & 0.074(17) & 0.69(88) & 0.0357(72) & -0.04(37) \\ 
10.3923 & 0.057(74) & -2.6(39) & 0.033(30) & 0.7(17) \\ 
\hline
\end{tabular}
\caption{Bare vacuum correlators for two different lattice spacings.
See the main text for the origin of the normalization factors and their 
uncertainties.}
\la{tab:T=0}
}
\TABLE[!p]{
\begin{tabular}{|c|r@{~~~~}r|r@{~~~~}r|}
\cline{1-5}
  \multicolumn{1}{|c|}{ }    & \multicolumn{2}{c |}{$20^4,\,\beta=6.200$, $\frac{dg_0^{-2}}{d\log a}=-0.1120$ }  
&  \multicolumn{2}{c|}{$24^4,\,\beta=6.408$, $\frac{dg_0^{-2}}{d\log a}=-0.1226$} \\
  \multicolumn{1}{|c|}{ $r/a$}    & \multicolumn{2}{c |}{ $\chi_s=1.262$, $Z_q=1.595$}  
&  \multicolumn{2}{c|}{$\chi_s=1.148$, $Z_q=1.495$} \\
 & \multicolumn{1}{c}{ $r^8C^{\rm bare}_{\theta\theta}/d_A$}    
&   \multicolumn{1}{c|}{ $-10^4\cdot r^8C^{\rm bare}_{qq}/d_A$}  & \multicolumn{1}{c}{ $r^8C^{\rm bare}_{\theta\theta}/d_A$} 
 &   \multicolumn{1}{c|}{ $-10^4\cdot r^8C^{\rm bare}_{qq}/d_A$ } \\
\cline{2-5}
\hline
2.0000  & 0.00100712(16) & -0.0263605(84) & 0.000664435(84) & -0.0182749(46) \\ 
3.0000  & 0.0045234(23) & 0.13444(12) & 0.0028715(14) & 0.110829(72) \\ 
4.0000  & 0.011065(21) & 0.4172(11) & 0.006926(13) & 0.35415(70) \\ 
5.0000  & 0.01758(12) & 0.4933(64) & 0.010103(74) & 0.4509(41) \\ 
6.0000  & 0.02506(53) & 0.514(27) & 0.01367(32) & 0.530(17) \\ 
7.0000  & 0.0331(18) & 0.456(96) & 0.0187(11) & 0.626(60) \\ 
8.0000  & 0.0448(54) & 0.65(27) & 0.0240(31) & 0.75(18) \\ 
9.0000  & 0.081(16) & 1.71(77) & 0.0315(81) & 1.15(46) \\ 
10.0000 & 0.191(44) & 3.0(24) & 0.038(19) & 0.2(10) \\ 
\hline
2.8284  & 0.0045366(14) & 0.132315(61) & 0.00302558(77) & 0.110874(35) \\ 
4.2426  & 0.012839(25) & 0.4221(12) & 0.007867(15) & 0.36367(77) \\ 
5.6569  & 0.02218(23) & 0.491(13) & 0.01209(14) & 0.4683(77) \\ 
7.0711  & 0.0308(13) & 0.432(76) & 0.01688(82) & 0.588(45) \\ 
8.4853  & 0.0368(59) & -0.06(32) & 0.0210(37) & 0.93(20) \\ 
9.8995  & 0.067(21) & -0.7(11) & 0.022(12) & 1.72(68) \\ 
\hline
3.4641  & 0.0084346(62) & 0.31326(31) & 0.0055463(36) & 0.26299(20) \\ 
5.1962  & 0.01849(14) & 0.4490(77) & 0.010459(86) & 0.4218(49) \\ 
6.9282  & 0.0325(15) & 0.455(76) & 0.01813(86) & 0.540(50) \\ 
8.6602  & 0.0336(85) & 0.48(44) & 0.0286(51) & 1.07(29) \\ 
10.3923 & 0.010(37) & -0.0(19) & 0.045(22) & 0.4(12) \\ 
\hline
\end{tabular}
\caption{Bare vacuum correlators for two different lattice spacings. }
\la{tab:T=0b}
}


\TABLE[!p]{
\begin{tabular}{|c|r@{~~~~}r|r@{~~~~}r|}
\cline{1-5}
  \multicolumn{1}{|c|}{ $r/a$}    & \multicolumn{2}{c |}{$6\times28^3$}  &  \multicolumn{2}{c|}{$9\times28^3$} \\
 & \multicolumn{1}{c}{ $r^8C^{\rm bare}_{\theta\theta}/d_A$}    
&   \multicolumn{1}{c|}{ $-10^4\cdot r^8C^{\rm bare}_{qq}/d_A$}  & \multicolumn{1}{c}{ $r^8C^{\rm bare}_{\theta\theta}/d_A$} 
 &   \multicolumn{1}{c|}{ $-10^4\cdot r^8C^{\rm bare}_{qq}/d_A$ } \\
\cline{2-5}
\hline
2.0000  & 0.00079523(22) & -0.0194011(81) & 0.00082743(11) & -0.0206983(56) \\ 
3.0000  & 0.0033813(42) & 0.13059(20) & 0.0035946(19) & 0.132683(96) \\ 
4.0000  & 0.007945(38) & 0.4061(20) & 0.008540(18) & 0.41965(93) \\ 
5.0000  & 0.01161(22) & 0.563(12) & 0.01270(10) & 0.5784(56) \\ 
6.0000  & 0.01689(93) & 0.696(52) & 0.01814(45) & 0.754(23) \\ 
7.0000  & 0.0245(33) & 0.69(17) & 0.0263(16) & 0.960(80) \\ 
8.0000  & 0.0335(100) & 0.17(53) & 0.0312(46) & 1.24(24) \\ 
9.0000  & 0.050(25) & -1.4(13) & 0.022(12) & 1.45(62) \\ 
10.0000 & 0.033(57) & 0.5(31) & -0.012(28) & 1.7(14) \\ 
\hline
2.8284  & 0.0035174(23) & 0.128916(88) & 0.0037160(11) & 0.129810(44) \\ 
4.2426  & 0.008844(45) & 0.4285(22) & 0.009647(21) & 0.4365(10) \\ 
5.6569  & 0.01319(43) & 0.675(23) & 0.01524(20) & 0.655(11) \\ 
7.0711  & 0.0212(26) & 0.94(14) & 0.0249(12) & 0.949(62) \\ 
8.4853  & 0.029(11) & 1.59(56) & 0.0356(52) & 0.92(28) \\ 
9.8995  & 0.018(38) & 1.9(19) & 0.036(18) & 1.76(92) \\ 
\hline
3.4641  & 0.006300(11) & 0.30305(55) & 0.0067527(53) & 0.30710(26) \\ 
5.1962  & 0.01114(27) & 0.563(14) & 0.01264(12) & 0.5582(66) \\ 
6.9282  & 0.0192(26) & 0.86(14) & 0.0219(12) & 0.871(65) \\ 
8.6602  & 0.004(16) & 1.30(85) & 0.0243(73) & 1.32(39) \\ 
10.3923 & -0.104(72) & 0.10(34) & 0.000(30) & -1.2(17) \\ 
\hline
\end{tabular}
\caption{Bare finite-temperature correlation functions at $\beta=6.2822$
($\chi_s=1.220$, $Z_q=1.546$, $\frac{dg_0^{-2}}{d\log a}=-0.1166$).}
\la{tab:6.2822}
}

\TABLE[!p]{
\begin{tabular}{|c|r@{~~~~}r|r@{~~~~}r|}
\cline{1-5}
\multicolumn{1}{|c|}{ $r/a$} & \multicolumn{2}{c |}{$6\times24^3$}  &  \multicolumn{2}{c|}{$8\times28^3$} \\
 & \multicolumn{1}{c}{ $r^8C^{\rm bare}_{\theta\theta}/d_A$}    
&   \multicolumn{1}{c|}{ $-10^4\cdot r^8C^{\rm bare}_{qq}/d_A$}  & \multicolumn{1}{c}{ $r^8C^{\rm bare}_{\theta\theta}/d_A$} 
 &   \multicolumn{1}{c|}{ $-10^4\cdot r^8C^{\rm bare}_{qq}/d_A$ } \\
\cline{2-5}
\hline
2.0000  & 0.000628574(92) & -0.0166504(39) & 0.000644215(92) & -0.0171814(40) \\ 
3.0000  & 0.0026309(18) & 0.110207(93) & 0.0027312(17) & 0.112619(90) \\ 
4.0000  & 0.006237(17) & 0.34772(93) & 0.006413(16) & 0.35577(89) \\ 
5.0000  & 0.008967(97) & 0.4805(55) & 0.008905(92) & 0.4822(54) \\ 
6.0000  & 0.01195(42) & 0.623(24) & 0.01185(41) & 0.618(23) \\ 
7.0000  & 0.0131(15) & 0.815(80) & 0.0156(14) & 0.747(76) \\ 
8.0000  & 0.0120(43) & 1.15(24) & 0.0217(40) & 0.83(23) \\ 
9.0000  & -0.001(11) & 1.87(61) & 0.028(10) & 0.60(60) \\ 
10.0000 & -0.042(25) & 3.8(14) & 0.014(25) & 0.4(14) \\ 
\hline
2.8284  & 0.00280374(100) & 0.111115(39) & 0.00290514(97) & 0.112818(40) \\ 
4.2426  & 0.006874(20) & 0.36775(100) & 0.007135(19) & 0.3723(10) \\ 
5.6569  & 0.00990(19) & 0.546(11) & 0.00984(18) & 0.5514(97) \\ 
7.0711  & 0.0139(12) & 0.811(64) & 0.0134(11) & 0.749(61) \\ 
8.4853  & 0.0190(48) & 1.03(27) & 0.0183(46) & 0.71(27) \\ 
9.8995  & 0.006(16) & 1.71(92) & 0.033(16) & 0.06(90) \\ 
\hline
3.4641  & 0.0050233(48) & 0.26228(25) & 0.0052397(47) & 0.26601(24) \\ 
5.1962  & 0.00836(12) & 0.4586(65) & 0.00886(11) & 0.4630(63) \\ 
6.9282  & 0.0136(12) & 0.814(65) & 0.0130(11) & 0.714(63) \\ 
8.6602  & 0.0133(68) & 0.64(39) & 0.0197(65) & 0.61(37) \\ 
10.3923 & -0.030(29) & 0.1(17) & 0.087(29) & -3.6(17) \\ 
\hline
\end{tabular}
\caption{Bare finite-temperature correlation functions at $\beta=6.408$
($\chi_s=1.148$, $Z_q=1.495$, $\frac{dg_0^{-2}}{d\log a}=-0.1226$).}
\la{tab:6.408}
}

\end{document}